\input phyzzx
\input epsf
\tolerance=10000
\sequentialequations
\def\rl{\rightline}
\def\ll{\leftline}

\def\etal{{\it et. al.}}
\def\r#1{$\bf#1$}

\def\t1{{\tilde 1}}

\def\AEF{A.E. Faraggi}
\def\DVN{D.V. Nanopoulos}

\def\SSM{supersymmetric standard model}
\def\NPB#1#2#3{Nucl. Phys. B{\bf#1} (19#2) #3}
\def\PLB#1#2#3{Phys. Lett. B{\bf#1} (19#2) #3}
\def\PRD#1#2#3{Phys. Rev. D{\bf#1} (19#2) #3}
\def\PRL#1#2#3{Phys. Rev. Lett. {\bf#1} (19#2) #3}
\def\PRT#1#2#3{Phys. Rep. {\bf#1} (19#2) #3}

\def\IJMP#1#2#3{Int. J. Mod. Phys. A{\bf#1} (19#2) #3}
\def\l{\langle}
\def\r{\rangle}
\tolerance=1000

\REF\GSW{For a review see, M. Green, J. Schwarz and E. Witten,
Superstring Theory, 2 vols., Cambridge University Press, 1987.}
\REF\heterotic{D. Gross, J. Harvey, E. Martinec and R. Rohm,
\PRL{54}{85}{502}; \NPB{256}{85}{253}.}
\REF\CHSW{P. Candelas, G. Horowitz, A. Strominger and E. Witten,
\NPB{258}{85}{46}.}
\REF\sy{L.J. Dixon, D. Friedan, E. Martinec and S. Shenker,
                                        \NPB{282}{87}{13};
M. Cvetic, \PRL{55}{87}{1795};
\PRL{59}{87}{2829}; \PRL{37}{87}{2366};
D. L\"ust, S. Theisen and G. Zoupanos, \NPB{296}{88}{800}.}
\REF\KLN{S. Kalara, J.L. Lopez and D.V. Nanopoulos,
                                        \PLB{245}{90}{421};
                                        \NPB{353}{91}{650}.}
\REF\wittenone{E. Witten, \PLB{155}{85}{151}; S. Ferrara, C. Kounnas and
M. Porrati, \PLB{181}{86}{263}.}
\REF\rsusy{For reviews, see:
                        H.P. Nilles, \PRT{110}{84}{1};
                        R. Arnowitt and P. Nath, {\it Applied N=1
                        Supergravity}\/ (World Scientific, Singapore, 1983);
                        H.E. Haber and G. L. Kane, \PRT{117}{85}{75};
                        D.V. Nanopoulos and A.B. Lahanas, \PRT{145}{87}{1}}
\REF\stringguts{D.C. Lewellen, \NPB{337}{90}{61};
     J. Ellis, J.L. Lopez and D.V. Nanopoulos, \PLB{245}{90}{375};
     A. Font, L.E. Ib\'a\~nez, and F. Quevedo, \NPB{345}{90}{389};
     S. Chaudhuri, S.-W. Chung, G. Hockney, and J. Lykken, 
	\NPB{456}{95}{89}, hep-ph/9501361;
     G. Aldazabal, A. Font, L.E. Ib\'a\~nez, and A.M. Uranga, 
	\NPB{452}{95}{3}, hep-th/9410206;
     G. Cleaver, hep-th/9506006;
     D. Finnell, \PRD{53}{96}{5781}, hep-th/9508073.}
\REF\SUTHREE{   M. Dine {\it{el al.}}, \NPB{259}{85}{549};
                B. Greene {\it{el al.}}, \PLB{180}{86}{69};
                \NPB{278}{86}{667}; {\bf B292} (1987) 606;
                R. Arnowitt and  P. Nath,
                \PRD{39}{89}{2006}; {\bf D42} (1990) 2498;
                Phys. Rev. Lett. {\bf 62} (1989) 222.}
\REF\REVAMP{I. Antoniadis, J. Ellis, J. Hagelin and D.V. Nanopoulos,
                \PLB{231}{89}{65}.}
\REF\ALR{I. Antoniadis, G.K. Leontaris and J. Rizos, \PLB{245}{90}{161};
                G.K. Leontaris, \PLB{372}{96}{212}, hep-ph/9601337.}
\REF\FSU{T.T. Burwick, A.K. Kaiser and H.F. Muller \NPB{362}{91}{232};
         A. Kagan and S. Samuel, \PLB{284}{92}{89}.}
\REF\SSM{L.E. Iba{\~n}ez {\it{et al.}}, \PLB{191}{87}{282};
        A. Font {\it{et al.}}, \PLB{210}{88}{101};
        A. Font {\it{et al.}}, Nucl.Phys. {\bf B331} (1990) 421;
        D. Bailin, A. Love and S. Thomas, \PLB{194}{87}{385};
                                          \NPB{298}{88}{75};
        J.A. Casas, E.K. Katehou and C. Mu{\~n}oz, \NPB{317}{89}{171};
        S. Chaudhuri, G. Hockney, and J. Lykken, 
		\NPB{461}{96}{357}, hep-th/9510241.}
\REF\FNY{\AEF, D.V. Nanopoulos and K. Yuan, \NPB{335}{90}{347}.}
\REF\EU{\AEF, \PLB{278}{92}{131}.}
\REF\TOP{\AEF, \PLB{274}{92}{47}.}
\REF\SLM{\AEF, \NPB{387}{92}{239}, hep-th/9208024.}
\REF\FFF{
        H. Kawai, D.C. Lewellen, and S.H.-H. Tye, \NPB{288}{87}{1};
        I. Antoniadis, C. Bachas, and C. Kounnas, \NPB{289}{87}{87};
        I. Antoniadis and C. Bachas, \NPB{298}{88}{586};
        R. Bluhm, L. Dolan, and P. Goddard, \NPB{309}{88}{330}.}
\REF\CDF{F. Abe \etal, \PRL{74}{95}{2626}, hep-ex/9503002;
           S. Abachi \etal, \PRL{74}{95}{2632}, hep-ex/9503003.}
\REF\DSW{M. Dine, N. Seiberg and E. Witten, Nucl. Phys.{\bf B289} (1987) 585.}
\REF\FI{P. Fayet, J. Iliopoulos, Phys. Lett.{\bf B51} (1974) 461.}
\REF\GCU{\AEF, \PLB{302}{92}{202}.}
\REF\DF{K.R. Dienes and \AEF, \PRL{75}{95}{2646};
                Nucl. Phys. {\bf B457} (1995) 409.}
\REF\DFM{ J.A. Casas and C. Munoz, \PLB{214}{88}{543};
   L. Ib\'a\~nez, \PLB{318}{93}{73};
   K.R. Dienes, \AEF, and J. March-Russell, 
		\NPB{467}{96}{44}, hep-th/9510223.}
\REF\tqmp{\AEF, \PLB{377}{96}{43}, hep-ph/9506388.}
\REF\LT{G.K. Leontaris and N.D. Tracas, \PLB{372}{96}{219}, hep-ph/9511280.}
\REF\naturalness{\AEF~ and D.V. Nanopoulos, \PRD{48}{93}{3288}.}
\REF\custodial{\AEF, \PLB{339}{94}{223}.}
\REF\bpz{A. Belavin, A. Polyakov and B. Zamolodchikov, \NPB{241}{84}{333}.}
\REF\fsz{P. Di Francesco, H. Saleur and J. Zuber, \NPB{290[FS20]}{87}{527}.}
\REF\YUKAWA{\AEF, \PRD{47}{93}{5021}.}
\REF\FM{\AEF, \NPB{407}{93}{57}.}
\REF\SFMM{J. Lopez and \DVN, \NPB{338}{90}{73}, \PLB{251}{90}73;
                                \PLB{256}{91}150; \PLB{268}{91}359;
          G.K. Leontaris, J. Rizos  and K. Tamvakis, \PLB{251}{90}{83};
          J. Rizos and K. Tamvakis, \PLB{251}{90}{369}.}
\REF\NRT{\AEF, \NPB{403}{93}{101}.}
\REF\rbtau{M. Chanowitz, J. Ellis and M. Gaillard, \NPB{128}{77}{506};
           A.J. Buras \etal, \NPB{135}{78}{66}.}
\REF\gs{D. Gross and J. Sloan, \NPB{291}{87}{41}.}
\REF\price{I. Antoniadis, J. Ellis, S. Kelley and D.V. Nanopoulos,
                \PLB{272}{91}{31}.}
\REF\ginsparg{P. Ginsparg, \PLB{197}{87}{139}.}
\REF\vadim{V. Kaplunovsky, \NPB{307}{88}{145};
                Erratum:  {\it ibid.}, {\bf B382}, 436 (1992).}
\REF\Gaillard{
      D. Bailin and A. Love, \PLB{280}{92}{26};
      M.K. Gaillard and R. Xiu, \PLB{296}{92}{71};
      J.L. Lopez, D. Nanopoulos, and K. Yuan,  \NPB{399}{93}{654};
      R. Xiu, \PRD{49}{94}{6656};
      S.P. Martin and P. Ramond, \PRD{51}{95}{6515}.}
\REF\scales{\AEF, \PRD{46}{92}{3204};
            \AEF~ and E. Halyo, \PLB{307}{93}{311};
            J.L. Lopez and D.V. Nanopoulos, 
			\PRL{76}{96}{1569}, hep-ph/9511426.}
\REF\BBO{V. Barger, M.S. Berger and P. Ohmann, \PRD{47}{93}{1093};
        H. Arason \etal,\PRD{46}{92}{3945}.}
\REF\expalpha{Particle Data Group, L. Montanet {\it et al,\/}
\PRD{50}{94}{1173} and 1995 off--year partial update for the
1996 edition available on the PDG WWW pages (URL: http://pdg.lbl.gov/)}
\REF\voloshin{For a recent analysis see for example, M.B. Voloshin,
             \IJMP{10}{95}{2865}, hep-ph/9502224.}
\REF\hrs{L. Hall, R. Rattazzi and U. Sarid, \PRD{50}{94}{7048}.}
\REF\mssmewx{see {\it e.g.}, the following papers and references therein:
                L.E. Ibanez and C. Lopez, \PLB{126}{83}{54};
        L. Alvarez-Gaum{\'e}, J. Polchinski and M. Wise, \NPB{221}{83}{495};
        G. Gamberini, G. Ridolfi and F. Zwirner, \NPB{331}{90}{331};
                M. Drees and M.M. Nojiri, \NPB{369}{92}{54};
                S. Kelley \etal, \NPB{398}{93}{3};
                P. Langacker and N. Polonsky, \PRD{50}{94}{2199}.}
\REF\cm{see {\it e.g.},
        J.M. Frere, D.R.T. Jones and S. Raby, \NPB{222}{83}{11};
  C. Kounnas, A.B. Lahanas, D.V. Nanopoulos and M. Quir{\'o}s,
        \NPB{236}{84}{438};
        J.A. Casas, A. Lleyda and Munoz, 
		\NPB{471}{96}{3}, hep-ph/9507294.}
\REF\FG{\AEF~and B. Grinstein, \NPB{422}{94}{3}.}
\REF\musolutions{see {\it e.g.},
        J.E. Kim and H.P. Nilles, \PLB{138}{84}{150};
        G.F. Giudice and A. Masiero, \PLB{206}{88}{480};
        J.A. Casas and C. Munoz, \PLB{306}{93}{288};
        I Antoniadis, E. Gava, K.S. Narain and T.R. Taylor, \NPB{432}{94}{187};
        Y.Nir, \PLB{354}{95}{107};
        V. Jain and R. Shrock, hep-ph/9507238. }
\REF\CKM{\AEF~and E. Halyo, \NPB{416}{94}{63}.}
\REF\yukthresh{See for example, B.D. Wright, hep-ph/9404217, and references
                                        therein.}
\REF\largetanbeta{M. Carena, M. Olechwski, S. Pokorski and C.E.M. Wagner,
                        \NPB{426}{94}{269}.}
\REF\PD{\AEF, \NPB{428}{94}{111}.}
\REF\ramond{ P. Binetruy and P. Ramond, \PLB{350}{95}{49}.}
\REF\LNZ{J.L. Lopez, D.V. Nanopoulos and A. Zichichi, \PRD{52}{95}{4178},
hep-ph/9502414.}
\REF\stcy{I. Antoniadis, E. Gava, K.S. Narain and T.R. Taylor,
\NPB{407}{93}{706}.}

\singlespace
\rl{IASSNS--HEP--95/60}
\rl{UFIFT--HEP--95-24}
\rl{hep-ph/9601332}
\nopagenumbers
\normalspace
\smallskip
\titlestyle{\bf Calculating Fermion masses in
                 Superstring Derived Standard--like Models}
\author{Alon E. Faraggi{\footnote*{
e--mail address: faraggi@phys.ufl.edu}}}
\smallskip
\centerline {Department of Physics, University of Florida}
\centerline {Gainesville, FL 33621}
\smallskip
\titlestyle{ABSTRACT}
\baselineskip=12pt
\smallskip
One of the intriguing achievements of the superstring derived
standard--like models in the free fermionic formulation
is the possible explanation of the
top quark mass hierarchy and the successful prediction of the top
quark mass. An important property of the superstring derived
standard--like models, which enhances their predictive power,
is the existence of three and only three generations in
the massless spectrum.
Up to some motivated assumptions with regard to the
light Higgs spectrum,
it is then possible to calculate
the fermion masses in terms of string tree level
amplitudes and some VEVs that parameterize the string vacuum.
I discuss the calculation of the
heavy generation masses in the superstring derived
standard--like models. The top quark Yukawa coupling
is obtained from a cubic level mass term while the
bottom quark and tau lepton mass terms are obtained from
nonrenormalizable terms. The calculation of the heavy fermion Yukawa
couplings is outlined in detail in a specific toy model.
The dependence of the effective bottom quark and tau lepton Yukawa
couplings on the flat directions at the string scale is examined.
The gauge and Yukawa couplings are extrapolated from the string
unification scale to low energies.
Agreement with $\alpha_{\rm strong}$, $\sin^2\theta_W$ and
$\alpha_{\rm em}$ at $M_Z$ is imposed, which necessitates the
existence of intermediate matter thresholds. The needed intermediate
matter thresholds exist in the specific toy model. The effect
of the intermediate matter thresholds on the extrapolated Yukawa
couplings in studied. It is observed that
the intermediate matter
thresholds also help to maintain the correct $b/\tau$ mass relation.
It is found that for a large portion of the
parameter space, the LEP precision data for
$\alpha_{\rm strong}$, $\sin^2\theta_W$ and
$\alpha_{\rm em}$, as well as the top quark mass
and the $b/\tau$ mass relation can all simultaneously be consistent
with the superstring derived standard--like models.
Possible corrections due to the supersymmetric mass spectrum are
studied as well as the minimization of the supersymmetric Higgs potential.
It is demonstrated that the calculated values of the Higgs VEV ratio,
$\tan\beta=v_1/v_2$, can be compatible with the minimization of the one--loop
effective Higgs potential.

\endpage

\pagenumbers
\pagenumber=1
\normalspace
\noindent{\bf 1. Introduction}
\smallskip
\baselineskip=16pt
One of the most important problems in elementary particle physics is
the origin of fermion masses. The Standard Model and its possible field
theoretic extensions, like Grand Unified Theories (GUTs) and supersymmetric
GUTs, do not provide means to calculate the fermion masses. In the context of
unified theories the fermion masses are expected to arise due to some
underlying Planck scale physics. Superstring theory [\GSW] is a unique
theory in the sense that it is believed to be a consistent theory of
quantum gravity while at the same time consistent heterotic string
vacua [\heterotic] give rise to massless spectra that closely resemble
the Standard Model [\CHSW]. At present, string theory provides the best
tool to probe Planck scale physics.

In the context of superstring theory one can
calculate the Yukawa couplings in terms of scattering amplitudes
between the string states and certain VEVs that parameterize the string
vacuum [\sy,\KLN]. In their low energy limit superstring theories give rise to
effective $N=1$ supergravity [\wittenone]. In the standard $N=1$ supergravity
model the electroweak Higgs VEV is fixed by the initial boundary conditions
at the unification scale and their evolution to the electroweak scale by the
renormalization group equations [\rsusy]. Thus, in superstring theories one
may be able to calculate the fermion masses. For this purpose one
must construct realistic superstring models. The construction of
realistic superstring models can be pursued in several approaches.
One possibility is to go through a simple [\stringguts] or a semi--simple
[\SUTHREE,\REVAMP,\ALR,\FSU] unifying group at intermediate energy scale.
Another is to derive the Standard Model directly from string
theory [\SSM,\FNY,\EU,\SLM,\TOP].

In Refs. [\EU,\TOP,\SLM] realistic
superstring standard--like models were constructed in the
four dimensional free fermionic formulation [\FFF].
One of the important achievements of the superstring derived standard--like
models in the free fermionic formulation is the possible
explanation of the top quark mass hierarchy and the successful prediction
of the top quark mass. In Ref. [\TOP] the top quark mass was predicted
to be in the approximate mass range $$m_t\approx175-180~GeV,\eqno(1)$$
three years prior to its experimental observation.
Remarkably, this prediction is in agreement with the top quark mass
as observed by the recent CDF and D0 collaborations [\CDF].

The superstring standard--like models have a very important property that
enhances their predictive power.
There are three and only three generations in the massless spectrum [\SLM].
There are no additional generations and mirror generations. Therefore,
the identification of the three light generations is unambiguous. This
property of the standard--like models enables, up to some
motivated assumptions with regard to the light Higgs spectrum,
unambiguous identification of the light fermion spectrum.
In this paper I will focus on the calculation of the heavy fermion masses.

The free fermionic standard--like models suggest an explanation for the
top quark mass hierarchy. At the cubic level of the superpotential
only the top quark gets a nonvanishing mass term. The mass terms for
the lighter quarks and leptons are obtained from nonrenormalizable
terms. Standard Model singlet fields in these
nonrenormalizable terms obtain nonvanishing
VEVs by the application of the Dine--Seiberg--Witten (DSW) mechanism [\DSW].
Thus, the order $N$ nonrenormalizable terms, of the form $cffh(\Phi/M)^{N-3}$,
become effective trilinear terms, where $f,h,\Phi$ denote fermions,
electroweak scalar doublets and Standard Model scalar singlets,
respectively. $M$ is a Planck scale mass to be defined later.
The effective Yukawa couplings are therefore given by
$\lambda=c(\langle \Phi \rangle/M)^{N-3}$.
The calculation of the coefficients $c$ for the heavy fermion family
is the main focus of the present paper.

In this paper I discuss the calculation of the heavy
fermion masses in the superstring derived standard--like models.
The analysis is illustrated in the toy model of Ref. [\TOP].
In this model the top quark Yukawa coupling is obtained
from a cubic level term in the superpotential while the bottom quark
and the tau lepton Yukawa couplings are obtained from quartic
order terms. The calculation of the cubic and quartic order correlators,
is described in detail. The Standard Model singlet fields
in the quartic order bottom quark and tau lepton mass terms
acquire a VEV by application of the DSW mechanism.
These VEVs parameterize the effective bottom quark
and tau lepton Yukawa couplings. The dependence of
the effective bottom quark and tau lepton Yukawa
couplings on the DSW VEVs is studied. It is shown that
there is substantial freedom in the resulting numerical
values of the effective Yukawa couplings. This freedom
in turn affects the low energy prediction of the top
quark mass.

The three heavy generation Yukawa couplings are extrapolated
from the unification scale to the electroweak scale by using
the coupled two--loop supersymmetric renormalization group
equations. Agreement with the low energy gauge parameters
$\alpha_{\rm em}(M_Z)$, $\sin^2\theta_W(M_Z)$ and
$\alpha_s(M_Z)$ is imposed. This requires that some
additional vector--like matter, beyond the MSSM and which
appear in the massless spectrum of the superstring standard--like
models, exist at intermediate energy scales [\GCU,\DF]. The
mass scales of the additional states is imposed by hand and
their derivation from the string model is left for future work.
The intermediate matter thresholds also affect the evolution of the
Yukawa couplings and consequently the low energy predictions of the
fermion masses [\tqmp,\LT].

The bottom quark and $W$--boson masses are used to calculate
the electroweak VEV ratio, $\tan\beta=v_1/v_2$. The extrapolated
Yukawa couplings and $\tan\beta$ are then used to calculate the
top quark mass and the ratio of the Yukawa couplings
$\lambda_b(M_Z)/\lambda_\tau(M_Z)$. As the VEV in the DSW mechanism,
which fixes the effective bottom quark and tau lepton Yukawa
couplings is varied, the predicted top quark mass
is found in the approximate range
$$90~{\rm GeV}\le m_t(m_t)\le205~{\rm GeV}.$$
and $\tan\beta$ is found in the approximate range
$$0.6\le\tan\beta\le28.$$

Thus, the predicted top quark mass can exist in a wide range
and is correlated with the predicted value of $\tan\beta$.
For fixed values of the VEVs in the DSW mechanism the
top quark mass and $\tan\beta$ are of course fixed.
The $b/\tau$ mass ratio is also found to be in good
agreement with experiment. It is found that
the intermediate matter thresholds which
are required for string gauge coupling unification
also help in maintaining the correct $b/\tau$ mass ratio.

In general, $\tan\beta$ can be fixed
by minimizing the Higgs potential. I examine the minimization of
the Higgs potential and illustrate that the calculated $\tan\beta$
can, in principle, be compatible with the minimization of the
one--loop Higgs effective potential. For this purpose, the soft
SUSY breaking parameters are fixed by hand and determination of
those terms in the string models is left for future work.

The paper is organized as follows. In section 2, I review the realistic
free fermionic models. Section  3 summarizes the tools needed for the
calculation of the Yukawa couplings. In section 4 the calculation
of the top quark Yukawa coupling is presented. In section 5 and 6
the calculation of the bottom quark and tau lepton Yukawa couplings
is described in detail. In section 5 the calculation of the quartic order
bottom quark and tau lepton mass terms is outlined.
In section 6 the dependence of the
effective bottom quark and tau lepton Yukawa couplings on the
DSW VEVs is investigated. In section 7 the top, bottom and tau
lepton Yukawa couplings are extrapolated to the electroweak scale,
in the presence the intermediate matter thresholds,
by using the coupled gauge and Yukawa two--loop RGEs. In section
8 I discuss the minimization of the one--loop Higgs
effective potential and possible corrections
from the supersymmetric mass spectrum. Section 9 concludes the paper.
\bigskip
\noindent{\bf 2. Realistic free fermionic models}
\smallskip
The free fermionic models are constructed by choosing a set of boundary
condition basis vectors and one--loop GSO projection coefficients [\FFF].
The possible boundary condition basis vectors and one-loop GSO phases
are constrained by the string consistency constraints. The physical states
are obtained by applying the generalized GSO projections. The physical
spectrum, its symmetries and interactions are then completely
determined. The low energy effective field theory is obtained
by $S$--matrix elements between external states. The Yukawa
couplings and higher order nonrenormalizable terms in the superpotential
are obtained by calculating correlators between vertex operators.
For a correlator to be nonvanishing all the symmetries of the model
must be conserved. Thus, the boundary condition basis vectors and
the one--loop GSO projection coefficients completely determine the
phenomenology of the models.

The first five basis vectors in the models that I discuss consist of the
NAHE set, $\{{\bf 1}, S,b_1,b_2,b_3\}$ [\naturalness,\SLM].
The vector $S$ in this set is the supersymmetry generator.
The two basis vectors $\{{\bf 1},S\}$ produce a model with $N=4$
space--time supersymmetry and $SO(44)$ gauge group.
At this level all of the internal world--sheet fermions are equivalent.
At the level of the NAHE set the gauge group is
$SO(10)\times SO(6)^3\times E_8$.
The sectors $b_1$, $b_2$ and $b_3$ each produce sixteen
spinorial 16 representation of $SO(10)$.
The number of generations is reduced to three and the $SO(10)$
gauge group is broken to one of its subgroups, $SU(5)\times U(1)$,
$SU(3)\times SU(2)\times U(1)^2$ or $SO(6)\times SO(4)$ by adding
to the NAHE set three additional basis vectors, $\{\alpha,\beta,\gamma\}$.
In the first two cases the basis vector that breaks the $SO(10)$
symmetry to $SU(5)\times U(1)$ must contain half integral boundary
conditions for the world--sheet complex fermions that generate the
$SO(10)$ symmetry. This basis vector is denoted as the vector $\gamma$.

The NAHE set plus the vector $2\gamma$ divide the world--sheet
fermions into several groups. The six left--moving real fermions,
$\chi^{1,\cdots,6}$ are paired to form three complex fermions
denoted $\chi^{12}$, $\chi^{34}$ and $\chi^{56}$. These complex
fermions produce the SUSY charges of the physical states. The sixteen
right--moving complex fermions $\bar\psi^{1\cdots5},\bar\eta^1,
\bar\eta^2,\bar\eta^3,\bar\phi^{1\cdots8}$ produce the observable
and hidden gauge groups, that arise from the sixteen dimensional
compactified space of the heterotic string in ten dimensions. The
complex world--sheet fermions, $\bar\psi^{1\cdots5}$, generate the
$SO(10)$ symmetry; $\bar\phi^{1,\cdots8}$ produce the hidden $E_8$
gauge group; and $\bar\eta^1$, $\bar\eta^2$, $\bar\eta^3$ give rise to
three horizontal $U(1)$ symmetries.
Finally, the twelve left--moving, $\{y,\omega\}^{1\cdots6}$, and
twelve right--moving, $\{{\bar y},\bar\omega\}^{1\cdots6}$, real fermions
correspond to the left/right symmetric internal conformal field
theory of the heterotic string, or equivalently to the six dimensional
compactified manifold in a bosonic formulation. The set of internal fermions
$\{y,\omega\vert{\bar y},{\bar\omega}\}^{1\cdots6}$ plays a fundamental
role in the determination of the low energy properties of the realistic
free fermionic models. In particular the assignment of boundary conditions,
in the vector $\gamma$, to this set of internal world--sheet fermions
selects cubic level Yukawa couplings for $+{2/3}$ or $-{1/3}$
charged quarks.

The three boundary condition basis vectors $\{\alpha,\beta,\gamma\}$
break the observable $SO(10)$ gauge group to one of its subgroups.
At the same time the horizontal symmetries are broken to factors of
$U(1)'s$. Three $U(1)$ symmetries arise from the complex right--moving
fermions $\bar\eta^1$, $\bar\eta^2$, $\bar\eta^3$. Additional horizontal
$U(1)$ symmetries arise by pairing two of the right--moving real
internal fermions $\{{\bar y},{\bar\omega}\}$. For
every right--moving $U(1)$ symmetry, there is a corresponding
left--moving global $U(1)$ symmetry that is obtained by pairing two of the
left--moving real fermions $\{y,\omega\}$. Each of the remaining
world--sheet left--moving real fermions from the set $\{y,\omega\}$ is paired
with a right--moving real fermion from the set
$\{{\bar y},{\bar\omega}\}$ to form a Ising model operator.

I now turn to describe the properties of the toy model of
Ref. [\TOP], which are important for the calculation
of the heavy fermion masses.
The three additional boundary condition basis vectors, beyond the NAHE
set, in the model of Ref. [\TOP] are given in table 1. In this toy model
an additional complication arises due to the appearance of additional
space--time vector bosons from twisted sectors [\custodial].
A combination of the $U(1)$ symmetries is enhanced to $SU(2)$.
The weak hypercharge then arises as a combination of the diagonal generator
of the custodial $SU(2)$ gauge group and the other $U(1)$ generators.
The custodial $SU(2)$ symmetry can be broken, near the Planck scale,
by a VEV of the custodial $SU(2)$ doublets, along F and D flat directions.
I will assume the existence of such a solution and neglect the effect
of the custodial $SU(2)$ symmetry. I will therefore focus on the
part of the gauge group that arises solely from the untwisted sector
and therefore on the properties that are
common to a large class of free fermionic models [\SLM].
The reason for illustrating the calculation in the toy model
of Ref. [\TOP] is because in this model nonvanishing bottom quark
and tau lepton mass terms arise at the quartic order of the
superpotential whereas, for example, in the model of Ref. [\EU]
such terms only appear at the quintic order.

In the models of Refs. [\EU,\TOP] the complex right--moving fermions
${\bar\psi}^{1,\cdots,5}$ produce the generators of the
$SU(3)\times SU(2)\times U(1)_C\times U(1)_L$ gauge group.
The right--moving complex fermions ${\bar\eta}^{1,2,3}$ generate three
$U(1)$ currents denoted by $U(1)_{r_{1,2,3}}$. Three additional
right--moving $U(1)$ symmetries, denoted $U(1)_{r_{j}}$ (j=4,5,6), arise
from three additional complexified right--moving fermions from the set
$\{{\bar y},{\bar\omega}\}$ denoted by
$$\eqalignno{
{\rm e}^{i{\bar\zeta}_{_1}}&={1\over\sqrt{2}}(\bar y^3+i\bar y^6),&(2a)\cr
{\rm e}^{i{\bar\zeta}_{_2}}&={1\over\sqrt2}(\bar y^1+i\bar\omega^5),&(2b)\cr
{\rm
e}^{i{\bar\zeta}_{_3}}&={1\over\sqrt2}(\bar\omega^2+i\bar\omega^4).&(2c)\cr}$$
For every local right--moving $U(1)_r$ symmetry there is
a corresponding global left--moving $U(1)_\ell$ symmetry.
The first three, denoted $U(1)_{\ell_j}$ $(j=1,2,3)$,
correspond to the charges of the supersymmetry generator
$\chi^{12}$, $\chi^{34}$ and $\chi^{56}$, respectively.
The last three,
denoted $U(1)_{\ell_{j}}$ $(j=4,5,6)$, arise
from the three additional complexified left--moving fermions from the set
$\{{y},{\omega}\}$ denoted by
$$\eqalignno{
{\rm e}^{i{\zeta}_1}&={1\over\sqrt{2}}(y^3+i y^6),&(3a)\cr
{\rm e}^{i{\zeta}_2}&={1\over\sqrt2}(y^1+i\omega^5),&(3b)\cr
{\rm e}^{i{\zeta}_3}&={1\over\sqrt2}(\omega^2+i\omega^4).&(3c)\cr}$$
Finally, in the models of Refs. [\EU, \TOP]
there are six Ising model operators
denoted by $$\sigma^i=\{\omega^1{\bar\omega}^1,
y^2{\bar y}^2, \omega^3{\bar\omega}^3, y^4{\bar y}^4,
y^5{\bar y}^5, \omega^6{\bar\omega}^6\},\eqno(4)$$
which are obtained by pairing a left--moving real fermion with a
right--moving real fermion.

The full massless spectrum of this model is given
in Ref. [\custodial]. Here I list only the states
that are relevant for the analysis of the heavy fermion
mass terms.
The sectors $b_1$, $b_2$ and $b_3$ produce three chiral generations,
$G_\alpha=e_{L_\alpha}^c+u_{L_\alpha}^c+N_{L_\alpha}^c+d_{L_\alpha}^c+
Q_\alpha+L_\alpha$ $(\alpha=1,\cdots,3)$, with charges under the
horizontal symmetries. For every generation, $G_j$ there are
two right--moving, $U(1)_{r_j}$ and $U(1)_{r_{j+3}}$, symmetries.
For every right--moving $U(1)$ gauged symmetry, there is a corresponding
left--moving global $U(1)$ symmetry, $U(1)_{\ell_j}$ and
$U(1)_{\ell_{j+3}}$. Each sector $b_1$, $b_2$ and $b_3$
has two Ising model operators,
($\sigma_4$, $\sigma_5$), ($\sigma_2$, $\sigma_6$) and
($\sigma_1$, $\sigma_3$),
respectively, obtained by pairing a left--handed real fermion with
a right--handed real fermion. In the superstring derived standard--like
models the vectors $b_1,b_2,b_3$ are the only vectors in the additive group
$\Xi$ which give rise to spinorial $16$ representation of $SO(10)$.
This property enhances the predictability of
the superstring derived standard--like models.

The Neveu--Schwarz (NS) sector corresponds to the untwisted sector of the
orbifold model and produces in addition to the gravity and gauge multiplets
three pairs of electroweak
scalar doublets $\{h_1, h_2, h_3, {\bar h}_1, {\bar h}_2, {\bar h}_3\}$,
three pairs of $SO(10)$ singlets with $U(1)$ charges,
$\{\Phi_{12},\Phi_{23},\Phi_{13},{\bar\Phi}_{12},
{\bar\Phi}_{23}, {\bar\Phi}_{13}\}$, and three singlets, which are singlets
of the entire four dimensional gauge group, $\xi_1,\xi_2,\xi_3$.

The sector ${S+b_1+b_2+\alpha+\beta}$ ($\alpha\beta$ sector) also
produces states that transform only under the observable gauge group.
In addition to two pairs of
electroweak doublets,
$\{h_{45},{\bar h}_{45},
h^\prime_{45},{\bar h}^\prime_{45}\}$,
there are four pairs of $SO(10)$
singlets with horizontal $U(1)$ charges,
$\{\Phi_{45},{\bar\Phi}_{45},\Phi^\prime_{45},{\bar\Phi}^\prime_{45},
\Phi_{1,2,},{\bar\Phi}_{1,2}\}$.

The spectrum described above is generic to a large class of
superstring standard--like models that utilize the NAHE set of basis
vectors. The states from the Neveu--Schwarz sector and the sectors
$b_1$, $b_2$ and $b_3$ are the states which arise from the underlying
$Z_2\times Z_2$ orbifold compactification. These states are therefore
common to all the superstring standard--like models that use the NAHE
set. Different models mainly differ by the assignment of boundary conditions
to the set of internal fermions $\{y,\omega\vert{\bar y},{\bar\omega}\}$
in the basis vectors beyond the NAHE set.
Consequently, the observable spectrum in different models differs by the
values of horizontal charges. A vector combination of the form
$b_1+b_2+\alpha+\beta$ is also common in the free fermionic models
that use the NAHE set. The states from this sector are important
in the free fermionic standard--like models for generating the fermion
mass hierarchy and for producing flat directions. Therefore, the
results discussed in this paper are shared by a large class of free
fermionic standard--like models.
\vfill
\eject
\bigskip
\noindent{\bf 3. Tools for calculating the fermion mass terms}
\smallskip
Here I summarize the well known tools needed for the analysis of the 
nonrenormalizable terms. Further details on the derivation of these rules
are given in ref. [\KLN]. 
Renormalizable and nonrenormalizable
contributions to the superpotential are obtained
by calculating correlators between vertex operators
$$A_N\sim\langle V_1^fV_2^fV_3^b\cdot\cdot\cdot V_N^b\rangle,\eqno(5)$$
where $V_i^f$ $(V_i^b)$ are the fermionic (scalar)
components of the vertex operators.
The vertex operators that appear in the fermion mass terms
have the following generic form,
$$\eqalignno{
V^{\ell}_{(q)}={\rm e}^{(qc)}~{\cal L}^\ell~
                                &{\rm e}^{(i\alpha\chi_{_{12}})}~
                                {\rm e}^{(i\beta\chi_{_{34}})}~
                                {\rm e}^{(i\gamma\chi_{_{56}})}~\cr
           &\left(~{\prod_{j}}{\rm e}^{(iq_i\zeta_{j})}~
                                                ~\{\sigma's\}~
                   {\prod_{j}}{\rm e}^{(i{\bar q}_i{\bar\zeta}_{j})}~\right)\cr
                                &{\rm e}^{(i{\bar\alpha}{\bar\eta}_{1})}~
                                {\rm e}^{(i{\bar\beta}{\bar\eta}_{2})}~
                                {\rm e}^{(i{\bar\gamma}{\bar\eta}_{3})}~
                                {\rm e}^{(iW_R\cdot{\bar J})}\cr
                        &{\rm e}^{(i{1\over2}KX)}~
                        {\rm e}^{({\rm i}{1\over2}K\cdot{\bar X})}&(6)}$$
where,
\parindent=-15pt
\item{\bullet} ${\rm e}^{(qc)}$ is the ghost charge, with conformal dimension
$$h=-{q^2\over2}-q.\eqno(7)$$ In the canonical picture $q=-{1/2}$ for fermions
and $q=-1$ for bosons.
\item{\bullet} ${\cal L}^\ell$ is the Lorentz group factor and signals
the space--time spin of a state. The space--time spin of a state is
determined by the boundary condition of the world--sheet
$\psi^\mu$ field. A periodic ${\psi^\mu}$ produces the
spinor representation of the Lorentz group and is represented
by the conformal field $S_\alpha$, where $\alpha$ is the space--time
spinor index. An antiperiodic ${\psi^\mu}$ produces space--time bosons,
denoted $\psi^\mu$ for vectors and $I$ for scalars. The conformal dimensions
of these fields are,
$$\eqalignno{
I~~~~~&(0,0)&(8a)\cr
S_\alpha~~~~~&({1\over4},0)&(8b)\cr
\psi^\mu~~~~~&({1\over2},0)&(8c)}$$
respectively.
\item{\bullet} ${\rm e}^{iqf}$ and ${\rm e}^{i{\bar q}{\bar f}}$ are the
factors that
arise from complexified fermions, which produce
global left--moving and local right--moving $U(1)$ currents, respectively.
A pair of left--moving (or right--moving) real fermions
$f_1$, $f_2$ which are complexified,
$$  f={1\over\sqrt2}(f_1+if_2)={\rm e}^{-iH}~~~~~~~,~~~~~~~
  f^*={1\over\sqrt2}(f_1-if_2)={\rm e}^{iH}\eqno(9)$$
produce a $U(1)$ current with charges,
$$Q(f)={1\over2}\alpha(f)+F(f),\eqno(10)$$
where $\alpha(f)$ and $F(f)$ are the boundary condition and fermion number
of the complex world--sheet fermion $f$. The conformal dimension of a
complex fermion is given by $h={q^2/2}$ and
${\bar h}={{\bar q}^2/2}$.
\item{\bullet} $\sigma's$: A left--moving real fermion, $f$, which is paired
with a right--moving real fermion ${\bar f}$, produces an Ising model operator
with the following conformal fields,
$$\eqalignno{
I~~~~~&(0,0),&(11a)\cr
\sigma_{\pm}(z,{\bar z})~~~~~&({1\over{16}},{1\over{16}}),&(11b)\cr
f(z)~~~~~&({1\over2},0),&(11c)\cr
f({\bar z})~~~~~&(0,{1\over2}),&(11d)\cr
\epsilon\equiv f(z)f({\bar z})~~~~~&({1\over2},{1\over2}),&(11e)\cr
}$$
where $\sigma_{\pm}$ are the order and disorder operators and $\epsilon$
is the energy operator.
The order and disorder operators arise when
both $f$ and ${\bar f}$ are periodic in a given sector $\alpha$.
The remaining fields arise when none, left or right, or both left
and right fermion oscillators act on the vacuum.
\item{\bullet} ${\rm e}^{(iW_R\cdot{\bar J})}$ is the factor that arises
due to the right--moving non--Abelian gauge group. The conformal
dimension is given by ${\bar h}={{W\cdot W}/2}$ where $W$ is the
weight vector of a representation $R$.
\item{\bullet} ${\rm e}^{(i{1\over2}KX)}$ and
               ${\rm e}^{({\rm i}{1\over2}K\cdot{\bar X})}$
arise from the Poincare quantum numbers.
\parindent=15pt
\smallskip

For the massless states, the conformal dimension $h=1$ and ${\bar h}=1$.
An important check on the normalization of the various $U(1)$ factors
is that indeed $h=1$ and ${\bar h}=1$ for the vertex operators of the
massless states.

The first step in calculating the fermion masses is extracting the possible
non vanishing correlators. This is achieved by imposing invariance under
all the local Abelian and non Abelian local gauge symmetries and the
other string selection rules that will be discussed below.
In order to verify that a potential order $N$ mass term is indeed
nonvanishing and to extract quantitative results from the string derived
models one must calculate the order $N$ correlators.
The second step is therefore the actual calculation of the potentially
nonvanishing correlators.

\smallskip
The tri--level string amplitude is given by
$$A_N={g^{N-2}\over{(2\pi)^{N-3}}}{\cal N}\int\prod_{i=1}^{N-3}d^2z_i
\langle{V_1^f(z_\infty)V_2^f(1)V_3^b(z_1)\cdots
V_{N_1}^b(z_{N-3})V_N^b(0)}\rangle,\eqno(12)$$
where ${\cal N}=\sqrt2$ is a normalization factor and
SL(2,C) invariance is used to fix the location of three
of the vertex operators at
$z=z_\infty,1,0$. For a correlator to be nonvanishing all the
symmetries of the model
must be conserved. Also for tree--level amplitudes the total
ghost charge must be $-2$.
Since a bosonic (fermionic) vertex operator in the canonical
picture carries ghost
charge $-1~(-1/2)$, picture changing is required for $N\ge4$ amplitudes.
To obtain the correct ghost charge some of the
vertex operators are picture
changed by taking
$$V_{q+1}(z)=\lim_{w\to z}
{\rm e}^{c}(w)T_F(w)V_{q}(z),\eqno(13)$$
where $T_F$ is the super current and in the
fermionic construction is given by
$$T_F=\psi^\mu\partial_\mu X+i{\sum_{I=1}^6}\chi_{_I}{y_{_I}}
\omega_{_I}=T_F^0+T_F^{-1}+T_F^{+1}\eqno(14)$$
with
$$T_F^{-1}={\rm e}^{-i\chi^{12}}\tau_{_{12}}+{\rm e}^{-i\chi^{34}}\tau_{_{34}}
+{\rm e}^{-i\chi^{56}}\tau_{_{56}}{\hskip .5cm};
{\hskip .5cm}T_F^{-1}=(T_F^{+1})^*\eqno(15)$$
where
$$\tau_{_{ij}}={i\over{\sqrt2}}(y^i\omega^i+iy^j\omega^j)\eqno(16)$$
and $${\rm e}^{\chi^{ij}}={1\over\sqrt{2}}(\chi^i+i\chi^j).\eqno(17)$$

In the models of Refs. [\EU,\TOP]
the complexified left--moving fermions are
$y^1\omega^5$, $\omega^2\omega^4$ and $y^3y^6$.
Thus, one of the fermionic states in every term
$y^i\omega^i$ $(i=1,...,6)$ is
complexified and therefore can be written,
for example for $y^3$ and $y^6$,
as
$$y^3={1\over{\sqrt2}}({\rm e}^{i\zeta_{_1}}+{\rm e}^{-i\zeta_{_1}})~~~,~~~
  y^6={1\over{{\sqrt2}i}}({\rm e}^{i\zeta_{_1}}-{\rm e}^{-i\zeta_{_1}}).
\eqno(18)$$
Consequently, every picture changing operation changes the
total
$U(1)_\ell=U(1)_{\ell_4}+U(1)_{\ell_5}+U(1)_{\ell_6}$ charge
by $\pm1$. An odd (even)  order term
requires an even (odd) number of picture changing
operations to get the correct ghost number [\KLN].
Thus, for $A_N$ to be non vanishing,
the total $U(1)_\ell$ charge, before picture
changing, has to be an odd (even)
number, for even (odd) order terms, respectively.
Similarly, in every pair $~y_i\omega_i~$, one real
fermion, either $y_i$ or
$\omega_i$, remains real and is paired with the corresponding
right--moving real fermion to produce an Ising model sigma operator.
Every picture changing operation changes the number of left--moving
real fermions by one.
This property of the standard--like models significantly
reduces the number of
potential non vanishing terms.

The following Operator Product Expansions (OPEs) are used
in the evaluation of the fermion mass terms
\parindent=-15pt
\item{\bullet} Ghosts
$$\eqalignno{
\langle{
{\rm e}^{(-c/2)}(z_1)
{\rm e}^{(-c/2)}(z_2)
{\rm e}^{(-c)}(z_3)}\rangle&=
z_{12}^{-1/4}z_{13}^{-1/2}z_{23}^{-1/2}&(19)\cr}$$
\item{\bullet} Lorentz group
$$\langle{S_\alpha(z_1)S_\beta(z_2)}\rangle=
C_{\alpha\beta}z_{12}^{-1/2}\eqno(20)$$
\item{\bullet} Correlator of exponentials
$$\langle{\prod_j
{\rm e}^{{\rm i}{\vec\alpha}_j\cdot{\vec J}}}\rangle=
\prod_{i<j}(z_{ij})^{{\vec\alpha}_i\cdot{\vec\alpha}_j}\eqno(21)$$
\item{\bullet} Ising model correlators [\bpz,\fsz],
$$\eqalignno{
\langle f(z_1)\sigma^{\pm}(z_2)\rangle&=
                        {1\over\sqrt2}z_{12}^{-1/2}\sigma^\mp(z_1)&(22a)\cr
\langle\sigma^{\pm}(z_1)\sigma^{\pm}(z_2)\rangle&=
                                z_{12}^{-1/8}({\bar z}_{12})^{-1/8}&(22b)\cr
\langle\sigma^{+}(z_1)\sigma^{-}(z_2)f(z_3)\rangle&=
                        {1\over\sqrt2}z_{12}^{3/8}({\bar z}_{12})^{-1/8}
(z_{13}z_{23})^{-1/2}&(22c)\cr
\langle\sigma^{+}(z_1)\sigma^{-}(z_2){\bar f}(z_3)\rangle&=
{1\over\sqrt2}z_{12}^{-1/8}({\bar z}_{12})^{3/8}
({\bar z}_{13}{\bar z}_{23})^{-1/2}&(22d)\cr
\langle\sigma_{\pm}(z_\infty)\sigma_{\pm}(1)
\sigma_{\pm}(z)\sigma_{\pm}(0)\rangle&=
{1\over\sqrt2}\vert{z_\infty\vert^{-1/4}}
\vert{1-z}\vert^{-1/4}\vert{z}\vert^{-1/4}
(1+\vert{z}\vert+\vert{1-z}\vert)^{1/2}\cr&&(22e)}$$
\parindent=15pt
\smallskip
where $$z_{ij}=z_i-z_j.\eqno(23)$$
\bigskip
\ll{\bf 4. Calculation of the Top quark Yukawa coupling}
\smallskip
The superstring derived standard--like models suggest a superstring
mechanism which explains the suppression of the lighter quark and
lepton masses relative to the top quark mass. These models suggest
that only the top quark gets a nonvanishing cubic level mass term
while the lighter quarks and leptons get their mass terms from
nonrenormalizable terms which are suppressed relative to the
leading cubic level terms.

The assignment of boundary conditions in the basis vector $\gamma$
for the internal world--sheet fermions,
$\{y,\omega\vert{\bar y},{\bar\omega}\}$ selects a cubic level mass term
for $+{2/3}$ or $-{1/3}$ charged quarks. For each of the
sectors $b_1$, $b_2$ and $b_3$ the fermionic boundary conditions
selects the cubic level Yukawa couplings according to the difference,
$$\Delta_j=\vert\gamma(L_j)-\gamma(R_j)\vert=0,1\eqno(24)$$
where $\gamma(L_j)/\gamma(R_j)$ are the boundary conditions in the vector
$\gamma$ for the internal world--sheet fermions from the set
$\{y,\omega\vert{\bar y},{\bar\omega}\}$, that are periodic in the vector
$b_j$. If $$\Delta_j=1\eqno(25)$$
then a Yukawa coupling for the $+{2/3}$ charged
quark from the sector $b_j$ is nonzero and the Yukawa coupling for the
$-{1/3}$ charged quark vanishes. The opposite occurs if
$\Delta_j=0$. Thus, the states from each of the sectors $b_1$, $b_2$ and
$b_3$ can have a cubic level Yukawa coupling for the $+{2/3}$ or
$-{1/3}$ charged quark, but not for both. We can construct string models
in which both $+{2/3}$ and $-{1/3}$ charged quarks get a cubic
level mass term. The model of table 2 is an example of such a model.
In this model,
$$\eqalignno{
\Delta_1&=\vert\gamma(y^3y^6)-
\gamma({\bar y}^3{\bar y}^6)\vert=1,&(26a)\cr
\Delta_2&=\vert\gamma(y^1\omega^6)-
\gamma({\bar y}^1{\bar\omega}^6)\vert=0,&(26b)\cr
\Delta_3&=\vert\gamma(\omega^1\omega^3)-
               \gamma({\bar\omega}^1{\bar\omega}^3)\vert=0.&(26c)\cr}$$
Consequently, in this model there is a cubic level mass term
for the $+{2/3}$ charged quark from the sector $b_1$ and cubic level mass
terms for the $-1/3$ charged quark and for charged leptons from the
sectors $b_2$ and $b_3$.

We can also construct string models in which only
$+{2/3}$ charged quarks get a nonvanishing cubic level mass term.
The model of table 1 is an example of such a model.
In this model
$$\eqalignno{
\Delta_1&=\vert\gamma(y^3y^6)-\gamma({\bar y}^3{\bar y}^6)\vert=1,&(27a)\cr
\Delta_2&=\vert\gamma(y^1\omega^5)-\gamma({\bar y}^1
{\bar\omega}^5)\vert=1,&(27b)\cr
\Delta_3&=\vert\gamma(\omega^2\omega^4)-
               \gamma({\bar\omega}^2{\bar\omega}^4)\vert=1.&(27c)\cr}$$
Therefore, in this model $\Delta_1=\Delta_2=\Delta_3=1$ and cubic
level mass terms are obtained for the $+2/3$ charged quarks from the sectors
$b_1$, $b_2$ and $b_3$.

In Ref. [\YUKAWA] the superstring up/down
selection rule is proven by using the string consistency constraints and
Eq.\ (18) to show that either the $+{2/3}$ or the $-{1/3}$ mass term is
invariant under the $U(1)_j$ symmetry.

In the model of Ref. [\TOP] the following terms are obtained
in the observable sector at the cubic level of the superpotential
$$\eqalignno{W_3&=\{(
{u_{L_1}^c}Q_1{\bar h}_1+{N_{L_1}^c}L_1{\bar h}_1+
{u_{L_2}^c}Q_2{\bar h}_2+{N_{L_2}^c}L_2{\bar h}_2+
{u_{L_3}^c}Q_3{\bar h}_3+{N_{L_3}^c}L_3{\bar h}_3)\cr
&\qquad
+{{h_1}{\bar h}_2{\bar\Phi}_{12}}
+{h_1}{\bar h}_3{\bar\Phi}_{13}
+{h_2}{\bar h}_3{\bar\Phi}_{23}
+{\bar h}_1{h_2}{\Phi_{12}}
+{\bar h}_1{h_3}{\Phi_{13}}
+{\bar h}_2{h_3}{\Phi_{23}}\cr
&\qquad
+\Phi_{23}{\bar\Phi}_{13}{\Phi}_{12}
+{\bar\Phi}_{23}{\Phi}_{13}{\bar\Phi}_{12}
+{\bar\Phi}_{12}({\bar\Phi}_1{\bar\Phi}_1
+{\bar\Phi}_2{\bar\Phi}_2)
+{{\Phi}_{12}}(\Phi_1\Phi_1
+\Phi_2\Phi_2)\cr
&\qquad
+{1\over2}\xi_3(\Phi_{45}{\bar\Phi}_{45}
+h_{45}{\bar h}_{45}+{\Phi_{45}^\prime}{{\bar\Phi}_{45}^\prime}
+h_{45}^\prime{\bar h}_{45}^\prime
+\Phi_1{\bar\Phi}_1+\Phi_2{\bar\Phi}_2)\cr
&\qquad
+h_3{\bar h}_{45}{\bar\Phi}_{45}^\prime
+{\bar h}_3h_{45}{\Phi}_{45}^\prime
+h_3{\bar h}_{45}^\prime\Phi_{45}
+{\bar h}_3h_{45}^\prime{\bar\Phi}_{45}\cr
&\qquad
+{1\over2}(\xi_1D_1{\bar D}_1+\xi_2D_2{\bar D}_2)
+{1\over\sqrt{2}}(D_1{\bar D}_2\phi_2+{\bar D}_1D_2{\bar\phi}_1)\}
,\quad&(28)\cr}$$

At the cubic level of the superpotential the $+{2/3}$ charged
quarks get nonvanishing mass terms,
 $${u_{L_1}^c}Q_1{\bar h}_1~+~
      {u_{L_2}^c}Q_2{\bar h}_2~+~
      {u_{L_3}^c}Q_3{\bar h}_3,\eqno(29)$$
while the $-{1/3}$ charged
quarks and the charged leptons cubic level mass terms vanish.
This selection mechanism results from the specific assignment
of boundary conditions that specify the string models, with $\Delta_j=1$ for
$(j=1,2,3)$. Any free fermionic standard--like model or flipped $SU(5)$
model (i.e. that uses the vector $\gamma$ with $1/2$ boundary conditions),
which satisfies the condition $\Delta_j=1$ for $(j=1,2,3)$ will
therefore have cubic level mass terms only for $+{2/3}$ charged quarks.

Due to the horizontal, $U(1)_{r_j}$, symmetries of the string models,
each of the chiral generations, from the sectors
$b_j$, $j=1,2,3$, can couple at the cubic level
only to one of the Higgs pairs $h_j$, $\bar h_j$.
This results due to the fact that the states from a sector
$b_j$ and the Higgs doublets $h_j$ and $\bar h_j$ are charged with respect
to one of the horizontal $U(1)_j$, $j=1,2,3$ symmetries.
Analysis of the renormalizable and nonrenormalizable Higgs mass terms suggests
that for some appropriate choices of flat F and D directions, only one pair of
the Higgs doublets remains light at low energies [\FM].
In the flipped $SU(5)$ string model and the
standard--like models, it has been found
that we must impose [\SFMM,\EU,\TOP,\NRT],
$$\langle{\Phi_{12},{\bar\Phi}_{12}}\rangle=0,\eqno(30)$$
and that $\Phi_{45}$, and ${\bar\Phi}_{13}$ or ${\bar\Phi}_{23}$,
must be different from zero. From this result and the cubic
level superpotential it follows that in any
flat F and D solution, $h_3$ and ${\bar h}_3$ obtain a Planck scale mass.
This result is a consequence of the symmetry of the vectors
$\alpha$ and $\beta$ with respect to the vectors $b_1$ and $b_2$.
At the level of the NAHE set there is a cyclic symmetry between
the sectors $b_1$, $b_2$ and $b_3$. The sectors $\alpha$ and $\beta$
break the cyclic symmetry.
The consequence is that $h_3$ and ${\bar h}_3$ do not contribute to
the light Higgs representations and obtain superheavy mass
from cubic level superpotential terms. At this level a residual $Z_2$
symmetry exist between the sectors $b_1$ and $b_2$ and is
broken further by the choices of flat directions.
Higher order nonrenormalizable terms then give superheavy mass to
${\bar h}_1$ or ${\bar h}_2$ [\NRT]. As a result only one
nonvanishing mass term, namely the top quark mass term, remains at low
energies. 
It should be emphasized that the detailed analysis of the 
Higgs mass spectrum in the superstring standard--like models 
was performed in the model of Ref. [\EU]. However, the observable 
massless spectrum in the models of Ref. [\TOP,\GCU] is similar to that
to the the model of Ref. [\EU], with slight variations in the 
charges under the horizontal charges. The models differ by the 
assignment of boundary conditions in the basis vectors 
$\{\alpha, \beta, \gamma\}$, which affects mainly the spectrum under the 
hidden sector and the horizontal charges. Consequently, it is expected
that similar results with regard to the light Higgs spectrum 
can be found in the models of ref. [\TOP,\GCU]. 
I therefore assume the existence of a solution with ${\bar h}_2$
as one of the light Higgs multiplets, in which case the top quark mass
term is $$u_2Q_2{\bar h}_2.$$

The coefficients of the cubic--level terms in the superpotential,
$\int d^2\theta\Phi_1\Phi_2\Phi_3$
are given by Eq.\ (12) with $N=3$,
$$A_3=g\sqrt{2}\langle{V_{1(-1/2)}^f(z_1)V_{2(-1/2)}^f(z_2)
V_{3(-1)}^b(z_3)}\rangle\eqno(31)$$
The vertex operators in the canonical picture in the top quark mass term,
$u_2Q_2{\bar h}_2$, are
$$\eqalignno{
{u^f_{2(-{1\over2})}}&=
{\rm e}^{(-{1\over2}c)}S_\alpha
{\rm e}^{({\rm i}{1\over2}\chi_{_{34}})}
{\rm e}^{({\rm i}{1\over2}\zeta_{_2})}
        \sigma^+_2\sigma^+_6
{\rm e}^{({\rm i}{1\over2}{\bar\zeta}_{_2})}
{\rm e}^{({\rm i}{1\over2}{\bar\eta}_{_2})}
{\rm e}^{({\rm i}{\bar J}_{16}\cdot{\bar W}_{16})}
{\rm e}^{({\rm i}{1\over2}KX)}
{\rm e}^{({\rm i}{1\over2}K{\bar X})},\cr
{Q^f_{2(-{1\over2})}}&=
{\rm e}^{(-{1\over2}c)} S_\beta
{\rm e}^{({\rm i}{1\over2}\chi_{_{34}})}
{\rm e}^{(-{\rm i}{1\over2}\zeta_{_2})}
        \sigma^+_2\sigma^+_6
{\rm e}^{(-{\rm i}{1\over2}{\bar\zeta}_{_2})}
{\rm e}^{({\rm i}{1\over2}{\bar\eta}_{_2})}
{\rm e}^{({\rm i}{\bar J}_{16}\cdot{\bar W}_{16})}
{\rm e}^{({\rm i}{1\over2}KX)}
{\rm e}^{({\rm i}{1\over2}K{\bar X})},\cr
{\bar h}_{2(-1)}^b&=
{\rm e}^{(-c)}
{\rm e}^{(-{\rm i}\chi_{_{34}})}
{\rm e}^{({\rm i}W_{10}\cdot{J_{10}})}
{\rm e}^{(-{\rm i}{\bar\eta}_{_2})}
{\rm e}^{({\rm i}{1\over2}KX)}
{\rm e}^{({\rm i}{1\over2}K{\bar X})},&(32)\cr
}$$
The cubic level amplitude is given by
$$\eqalignno{A_3=g\sqrt{2}\biggl\{\biggr.
&\langle{
{\rm e}^{(-c/2)}(z_1)
{\rm e}^{(-c/2)}(z_2)
{\rm e}^{(-c)}(z_3)}\rangle\cr
&\langle{S_\alpha(z_1)S_\beta(z_2)}\rangle\cr
&\langle{
{\rm e}^{( {\rm i}{1\over2}\chi_{_{34}})}(z_1)
{\rm e}^{( {\rm i}{1\over2}\chi_{_{34}})}(z_2)
{\rm e}^{(-{\rm i}\chi_{_{34}})}(z_3)}\rangle\cr
&\langle{
{\rm e}^{({\rm i}{1\over2}\zeta_{_2})}(z_1)
{\rm e}^{(-{\rm i}{1\over2}\zeta_{_2})}(z_2)}\rangle\cr
&\langle{\sigma^+_2(z_1)\sigma^+_2(z_2)}\rangle\cr
&\langle{\sigma^+_6(z_1)\sigma^+_6(z_2)}\rangle\cr
&\langle{
{\rm e}^{({\rm i}{1\over2}{\bar\zeta}_{_2})}(z_1)
{\rm e}^{(-{\rm i}{1\over2}{\bar\zeta}_{_2})}(z_2)}\rangle\cr
&\langle{
{\rm e}^{({\rm i}{\bar J}_{\bar3,1}\cdot{\bar W}_{\bar3,1})}(z_1)
{\rm e}^{({\rm i}{\bar J}_{3,2}\cdot{\bar W}_{3,2})}(z_2)
{\rm e}^{({\rm i}{\bar J}_{1,2}\cdot{\bar W}_{1,2})}(z_3)}\rangle\cr
&\langle{
{\rm e}^{({\rm i}{1\over2}{\bar\eta}_{_2})}(z_1)
{\rm e}^{({\rm i}{1\over2}{\bar\eta}_{_2})}(z_2)
{\rm e}^{(-{\rm i}{}{\bar\eta}_{_2})}(z_3)}\rangle\cr
&\left\langle{{\prod_{i=1}^4}
{\rm e}^{[{\rm i}{1\over2}K_iX(i)]}
{\rm e}^{[{\rm i}{1\over2}K_i{\bar X}(i)]}}
\right\rangle\biggl.\biggr\}~.&(33)\cr}$$
The correlators are evaluated using the formula given in Eqs.\ (19-23). Since
$K_1+K_2+K_3=0$ and $K_1^2=K_2^2=K_3^2=0$, it follows that
$K_1\cdot K_2=K_1\cdot K_3=K_2\cdot K_3$=0.
Consequently, evaluation of the correlator in Eq.\ (33) yields,
$$A_3=g\sqrt{2}\eqno(34)$$
which is taken as the top quark Yukawa coupling at the
string unification scale.
\bigskip
\ll{\bf 5. Calculation of the bottom quark and tau lepton mass terms}
\smallskip
In free fermionic standard--like (and flipped $SU(5)$) models
with $\Delta_{1,2,3}=1$, only $+2/3$ charged quarks obtain potential
mass terms at the cubic level of the superpotential. There are no
potential cubic level mass terms for $-1/3$ charged quarks and for
charged leptons. A realistic string model must give rise to
such mass terms. Consequently, in this class of models,
$-1/3$ charged quarks and charged
leptons must get their mass terms from nonrenormalizable terms in
the superpotential. The nonrenormalizable terms have the general
form
$$
cf_if_jh(\phi/M)^{n-3}\eqno(35)
$$
where $c$ are the calculable coefficients of the $n^{\rm th}$
order correlators,
$f_i$, $f_j$ are the quark and lepton fields, $h$ are the light
Higgs representations and $\phi$ are Standard Model singlets in the
massless spectrum of the string models.
The scale $M$ is related to the Planck scale, and numerically
$M\sim1.2\cdot10^{18}GeV$.

In the models of Ref. [\EU,\TOP], due to the up/down Yukawa
superstring selection mechanism there are no potential mass terms
for $-1/3$ charged quarks and for charged leptons at the cubic level
of the superpotential. Such mass terms may arise from quartic,
quintic or higher order terms in the superpotential. In the
model of Ref. [\EU], for example, because of the global
$U(1)_{\ell_{j+3}}$ symmetries, there are no potential
bottom quark and tau lepton mass terms at the quartic
order of the superpotential [\YUKAWA].
In this model, $Q_{\ell_{j+3}}(Q_j,L_j)=-Q_{\ell_{j+3}}(d_j,e_j).$
Consequently, the total $U_\ell$ charge before picture changing
vanishes and the quartic order down quark and tau lepton mass terms
vanish. In the model
of Ref. [\EU] such potential non vanishing mass terms arise
at the quintic order of the superpotential
(in the notation of Ref. [\EU]),
$$W_5=\{{d_{L_1}^c}Q_1h_{45}\Phi_1^-\xi_2
+{e_{L_1}^c}L_1h_{45}\Phi_1^+\xi_2+
{d_{L_2}^c}Q_2h_{45}{\Phi}_2^-\xi_1
+{e_{L_2}^c}L_2h_{45}{\bar\Phi}_2^-\xi_1\}.\eqno(36)$$
The evaluation of the coefficient of the quintic
order terms involves a two dimensional complex
integration. Considerable simplification will be provided if we
can construct a model in which potential non--vanishing
bottom quark and tau lepton mass terms are obtained
from quartic order terms. Such a model was
constructed in Ref. [\TOP]. In this model
$Q_{\ell_{j+3}}(Q_j,L_j)=+Q_{\ell_{j+3}}(d_j,e_j).$
Consequently, the total $U_\ell$ charge before picture changing
is $\pm1$. In the model of Ref. [\TOP], the following non vanishing
mass terms for $-{1/3}$ charged quarks and for charged
leptons are obtained at the quartic order,
$$W_4=\{{d_{L_1}^c}Q_1h_{45}^\prime\Phi_1+
        {e_{L_1}^c}L_1h_{45}^\prime\Phi_1+
        {d_{L_2}^c}Q_2h_{45}^\prime{\bar\Phi}_2+
        {e_{L_2}^c}L_2h_{45}'{\bar\Phi}_2\}.\eqno(37)$$
This quartic order terms can therefore be potential
mass terms for the bottom quark and tau lepton.
To evaluate the bottom quark and tau lepton masses
we must first evaluate the coefficients of the quartic
order correlators. The Standard Model singlet in the
quartic order terms can then get a VEV, which then
results in effective bottom quark and tau lepton
Yukawa couplings. From Eq.\ (37) it is seen that if
${\bar\Phi}_2>>{\Phi}_1$ then the last two terms
in Eq.\ (37) are the bottom quark and tau lepton
mass terms.

For the bottom quark mass term, the vertex operators
in the canonical picture are given by
$$\eqalignno{
{d^f_{2(-{1\over2})}}&=
{\rm e}^{(-{1\over2}c)}~S_\alpha
{}~{\rm e}^{( {\rm i}{1\over2}\chi_{_{34}})}
{}~{\rm e}^{(-{\rm i}{1\over2}\zeta_{_2})}
        ~\sigma^+_2\sigma^-_6
{}~{\rm e}^{({\rm i}{1\over2}{\bar\zeta}_{_2})}
{}~{\rm e}^{({\rm i}{1\over2}{\bar\eta}_{_2})}
{}~{\rm e}^{({\rm i}{\bar J}_{16}\cdot{\bar W}_{16})}
{}~{\rm e}^{({\rm i}{1\over2}KX)}
{\rm e}^{({\rm i}{1\over2}K{\bar X})},\cr
{Q^f_{2(-{1\over2})}}&=
{\rm e}^{(-{1\over2}c)}~S_\beta
{}~{\rm e}^{({\rm i}{1\over2}\chi_{_{34}})}
{}~{\rm e}^{(-{\rm i}{1\over2}\zeta_{_2})}
        ~\sigma^+_2\sigma^+_6
{}~{\rm e}^{(-{\rm i}{1\over2}{\bar\zeta}_{_2})}
{}~{\rm e}^{({\rm i}{1\over2}{\bar\eta}_{_2})}
{}~{\rm e}^{({\rm i}{\bar J}_{16}\cdot{\bar W}_{16})}
{}~{\rm e}^{({\rm i}{1\over2}KX)}
{\rm e}^{({\rm i}{1\over2}K{\bar X})},\cr
{h'^b_{45(-1)}}&=
{\rm e}^{(-c)}
{}~{\rm e}^{(-{\rm i}{1\over2}\chi_{_{12}})}
 {\rm e}^{(-{\rm i}{1\over2}\chi_{_{34}})}
        ~\sigma^+_1\sigma^+_2\sigma^+_3\sigma^+_4
{}~{\rm e}^{(-{\rm i}{1\over2}{\bar\eta}_{_1})}
 {\rm e}^{(-{\rm i}{1\over2}{\bar\eta}_{_2})}
{}~{\rm e}^{({\rm i}{\bar J}_{10}\cdot{\bar W}_{10})}
{}~{\rm e}^{({\rm i}{1\over2}KX)}
 {\rm e}^{({\rm i}{1\over2}K{\bar X})},\cr
 {{\bar\Phi}^b_{2(-1)}}&=
{\rm e}^{(-c)}
{}~{\rm e}^{(-{\rm i}{1\over2}\chi_{_{12}})}
 {\rm e}^{(-{\rm i}{1\over2}\chi_{_{34}})}
        ~\sigma^-_1\sigma^+_2\sigma^+_3\sigma^+_4{\bar\omega}^6
{}~{\rm e}^{({\rm i}{1\over2}{\bar\eta}_{_1})}
{}~{\rm e}^{(-{\rm i}{1\over2}{\bar\eta}_{_2})}
{}~{\rm e}^{({\rm i}{1\over2}KX)}
 {\rm e}^{({\rm i}{1\over2}K{\bar X})},\cr}$$
and similarly for the tau lepton mass term,
$$\eqalignno{
{e^f_{2(-{1\over2})}}&=
 {\rm e}^{(-{1\over2}c)}~S_\alpha
{}~{\rm e}^{({\rm i}{1\over2}\chi_{_{34}})}
{}~{\rm e}^{({\rm i}{1\over2}\zeta_{_2})}
        ~\sigma^+_2\sigma^+_6
{}~{\rm e}^{({\rm i}{1\over2}{\bar\zeta}_{_2})}
{}~{\rm e}^{({\rm i}{1\over2}{\bar\eta}_{_2})}
{}~{\rm e}^{({\rm i}{\bar J}_{16}\cdot{\bar W}_{16})}
{}~{\rm e}^{({\rm i}{1\over2}KX)}
 {\rm e}^{({\rm i}{1\over2}K{\bar X})},\cr
{L^f_{2(-{1\over2})}}&=
{\rm e}^{(-{1\over2}c)}~S_\beta
{}~{\rm e}^{({\rm i}{1\over2}\chi_{_{34}})}
{}~{\rm e}^{({\rm i}{1\over2}\zeta_{_2})}
        ~\sigma^+_2\sigma^-_6
{}~{\rm e}^{(-{\rm i}{1\over2}{\bar\zeta}_{_2})}
{}~{\rm e}^{({\rm i}{1\over2}{\bar\eta}_{_2})}
{}~{\rm e}^{({\rm i}{\bar J}_{16}\cdot{\bar W}_{16})}
{}~{\rm e}^{({\rm i}{1\over2}KX)}
 {\rm e}^{({\rm i}{1\over2}K{\bar X})},\cr
{h'^b_{45(-1)}}&=
{\rm e}^{(-c)}
{}~{\rm e}^{(-{\rm i}{1\over2}\chi_{_{12}})}
 {\rm e}^{(-{\rm i}{1\over2}\chi_{_{34}})}
        ~\sigma^+_1\sigma^+_2\sigma^+_3\sigma^+_4
{}~{\rm e}^{(-{\rm i}{1\over2}{\bar\eta}_{_1})}
 {\rm e}^{(-{\rm i}{1\over2}{\bar\eta}_{_2})}
{}~{\rm e}^{({\rm i}{\bar J}_{10}\cdot{\bar W}_{10})}
{}~{\rm e}^{({\rm i}{1\over2}KX)}
{\rm e}^{({\rm i}{1\over2}K{\bar X})},\cr
{{\bar\Phi}^b_{2(-1)}}&=
{\rm e}^{(-c)}
{}~{\rm e}^{(-{\rm i}{1\over2}\chi_{_{12}})}
 {\rm e}^{(-{\rm i}{1\over2}\chi_{_{34}})}
        \sigma^-_1\sigma^+_2\sigma^+_3\sigma^+_4{\bar\omega}^6
{}~{\rm e}^{({\rm i}{1\over2}{\bar\eta}_{_1})}
 {\rm e}^{(-{\rm i}{1\over2}{\bar\eta}_{_2})}
{}~{\rm e}^{({\rm i}{1\over2}KX)}
 {\rm e}^{({\rm i}{1\over2}K{\bar X})},\cr}$$

It is observed that in this toy model
$\lambda_b(M_{\rm string})=\lambda_\tau(M_{\rm string})$ and it will be
sufficient to calculate one of the two. This $SU(5)$ relation [\rbtau]
is a reflection
of the underlying $SO(10)$ symmetry at the level of the
NAHE set. As is evident from Eq.\ (36), such a residual symmetry
does not necessarily survive the $SO(10)$ symmetry breaking vectors beyond
the NAHE set. Other superstring standard--like models can therefore yield
$\lambda_b(M_{\rm string})\ne\lambda_\tau(M_{\rm string})$.

The picture changed vertex operator for the
${\bar\Phi}_{2(-1)}^b$ field is obtained from Eq.\ (13),
$${\bar\Phi}_{2(0)}(z)=\lim_{w\to z}
{\rm e}^{c}(w)T_F(w){\bar\Phi}_{2(-1)}(z). \eqno(38)$$

Using the OPE
$${\rm e}^{{\rm i}\alpha J}(w)
  {\rm e}^{ {\rm i}\beta  J}(z)\sim
  (w-z)^{\alpha\cdot\beta}~
  {\rm e}^{ {\rm i}(\alpha + \beta) J}$$
and Eq.\ (22a) we obtain
$$\eqalignno{
{{\bar\Phi}^b_{2(0)}}=
{{\rm i}\over2}{\left\{\right.}
 &{\rm e}^{( {\rm i}{1\over2}\chi_{_{12}})}
  {\rm e}^{(-{\rm i}{1\over2}\chi_{_{34}})}
   (y_1\sigma^+_1\sigma^+_2+{\rm i}\omega_2\sigma^-_1\sigma^-_2)
\sigma^+_3\sigma^+_4\cr
+
 &{\rm e}^{(-{\rm i}{1\over2}\chi_{_{12}})}
  {\rm e}^{( {\rm i}{1\over2}\chi_{_{34}})}
(y_3\sigma^-_3\sigma^+_4+{\rm i}\omega_4\sigma^+_3\sigma^-_4)
\sigma^-_1\sigma^+_2{\left.\right\}}\cr
&{\bar\omega}^6
{\rm e}^{({\rm i}{1\over2}{\bar\eta}_{_1})}
{\rm e}^{(-{\rm i}{1\over2}{\bar\eta}_{_2})}
{\rm e}^{({\rm i}{1\over2}KX)}
{\rm e}^{({\rm i}{1\over2}K{\bar X})}~.~~&(39)\cr}$$
Only the first term contributes to the nonvanishing quartic order correlator,
which is given by
$$\eqalignno{
A_4
&={g^2\over{2\pi}}\int d^2{\rm z}\cr
&\langle{
{\rm e}^{(-{1\over2}c)}(1)
{\rm e}^{(-{1\over2}c)}(2)
{\rm e}^{(-c)}(3)}\rangle\cr
&\langle{
{\rm e}^{({\rm i}{1\over2}\chi_{_{12}})}(3)
{\rm e}^{(-{\rm i}{1\over2}\chi_{_{12}})}(4)}\rangle\cr
&\langle{
{\rm e}^{({\rm i}{1\over2}\chi_{_{34}})}(1)
{\rm e}^{({\rm i}{1\over2}\chi_{_{34}})}(2)
{\rm e}^{(-{\rm i}{1\over2}\chi_{_{34}})}(3)
{\rm e}^{(-{\rm i}{1\over2}\chi_{_{34}})}(4)}\rangle\cr
&\langle{
{\rm e}^{(-{\rm i}{1\over2}\zeta_{_2})}(1)
{\rm e}^{(-{\rm i}{1\over2}\zeta_{_2})}(2)
{\rm e}^{({\rm i}\zeta_{_2})}}(4)\rangle\cr
&\langle{\rm sigmas}\rangle\cr
&\langle{
{\rm e}^{({\rm i}{1\over2}{\bar\zeta}_{_2})}(1)
{\rm e}^{(-{\rm i}{1\over2}{\bar\zeta}_{_2})}(2)}\rangle\cr
&\langle{
{\rm e}^{({\rm i}{\bar J}_{\bar3,1}\cdot{\bar W}_{\bar3,1})}(1)
{\rm e}^{({\rm i}{\bar J}_{3,2}\cdot{\bar W}_{3,2})}(2)
{\rm e}^{({\rm i}{\bar J}_{1,2}\cdot{\bar W}_{1,2})}(3)}\rangle\cr
&\langle{
{\rm e}^{(-{\rm i}{1\over2}{\bar\eta}_{_2})}(1)
{\rm e}^{(-{\rm i}{1\over2}{\bar\eta}_{_2})}(2)
{\rm e}^{({\rm i}{1\over2}{\bar\eta}_{_2})}(3)
{\rm e}^{({\rm i}{1\over2}{\bar\eta}_{_2})}(4)}\rangle\cr
&\left\langle{{\prod_{i=1}^4}
         {\rm e}^{[{\rm i}{1\over2}K_iX(i)]}
         {\rm e}^{[{\rm i}{1\over2}K_i{\bar X}(i)]}}\right\rangle&(40)\cr}$$
where
$$\eqalignno{
\langle{\rm sigmas}\rangle=
&\langle{\sigma^+_1(3)\sigma^+_1(4)}\rangle\cr
&\langle{\sigma^+_2(1)\sigma^+_2(2)\sigma^+_2(3)\sigma^+_2(4)}\rangle\cr
&\langle{\sigma^+_3(3)\sigma^+_3(4)}\rangle\cr
&\langle{\sigma^+_4(3)\sigma^+_4(4)}\rangle\cr
&\langle{\sigma^-_6(1)\sigma^+_6(2){\bar\omega}_6(4)}\rangle~.&(41)\cr}$$
The correlator is evaluated by using Eqs.\ (19--23). In addition
the correlator due to the kinetic quantum numbers yields,
$$\eqalignno{\left\langle{{\prod_{i=1}^4}
         {\rm e}^{[{\rm i}{1\over2}K_iX(i)]}
         {\rm e}^{[{\rm i}{1\over2}K_i{\bar X}(i)]}}\right\rangle
        &=\prod_{i<j}\vert{z_{ij}}\vert^{{1\over2}K(i)\cdot K(j)}\cr&=
        \vert{z_\infty}\vert^{-{1\over4}(s+t+u)}\vert{1-z}\vert^{-{u\over4}}
                        \vert{z}\vert^{-{s\over4}}&(42)\cr}$$
where $s$, $t$ and $u$ are the usual Mandelstam variables with,
$$\eqalignno{
s&=-2~K(1)\cdot K(2)~=~-2~K(3)\cdot K(4)\cr
t&=-2~K(1)\cdot K(3)~=~-2~K(2)\cdot K(4)\cr
u&=-2~K(1)\cdot K(4)~=~-2~K(2)\cdot K(3)&(43)\cr
}$$
and $(s+t+u)=0$.
All the Standard Model states from the NS sectors, the sectors $b_j$ (j=1,2,3),
and the sector $b_1+b_2+\alpha+\beta$ fall into representations of the
underlying SO(10) symmetry. We can use the corresponding weights under
the $SO(10)$ symmetry to evaluate the correlator under the $SO(10)$ subgroup.
Therefore, the correlator under the $SO(10)$ gauge group yields,
$$\eqalignno{
\langle{
{\rm e}^{({\rm i}{\bar J}_{\bar3,1}\cdot{\bar W}_{\bar3,1})}(z_1)
{\rm e}^{({\rm i}{\bar J}_{3,2}\cdot{\bar W}_{3,2})}(z_2)
{\rm e}^{({\rm i}{\bar J}_{1,2}\cdot{\bar W}_{1,2})}(z_3)}\rangle&=\cr
{\bar z}_{12}^{W_{16}\cdot{W_{16^\prime}}}{\bar z}_{13}^{W_{16}\cdot{W_{10}}}
&{\bar z}_{23}^{W_{16^\prime}\cdot{W_{10}}}C_{16\cdot16\cdot10}&(44a)\cr}$$
where
$$W_{16}\cdot{W_{16^\prime}}=-3/4~;
 ~W_{16}\cdot{W_{10}}=
  W_{16^\prime}\cdot{W_{10}}=-1/2\eqno(44b)$$
and
$$W_{16}+W_{16^\prime}+W_{10}=0~~~~;~~~~
  W_{16}^2=W_{16^\prime}^2=5/4~;~W_{10}^2=1\eqno(44c)$$
SL(2,C) invariance is used to fixed three of the points,
$z_1=\infty$, $z_2=1$, $z_3=z$ $z_4=0$.
Using the OPEs and collecting all the terms we obtain the one dimensional
complex integral,
$$I=\int{d^2z\vert{z}\vert^{-{s\over4}-1}\vert{1-z}\vert^{-{u\over4}-{7\over4}}
(1+\vert{z}\vert+\vert{1-z}\vert)^{1\over2}}.\eqno(45)$$
To obtain the contact term we set $s=u=0$.
The integral is then evaluated numerically by shifting
$z\rightarrow1-z$ and using polar
coordinates,
$$I=2\int_0^\infty dr\int_0^\pi d\theta~
     r^{-3/4}(1-2r\cos\theta+r^2)^{-1/2}
    \{1+r+\sqrt{1-2r\cos\theta+r^2}\}^{1/2}\approx77.7$$
and
$$A_4={{g^2}\over{2\pi}}{1\over4}I.\eqno(46)$$
In general, to determine the contact term at this stage,
one needs to subtract
the field theory contributions to the four point amplitude [\gs].
Possible graviton, gauge and massless matter fields
must be accounted for.
However, for the terms in Eq.\ (37), the charges of the
fields involved are such that
no graviton or gauge boson exchanges are possible.
This can be seen from superfield diagrams
for the gauge boson : $\Phi^\dagger{\rm e}^{gV}\Phi$,
and the fact that $\Phi^\dagger$
cannot appear in the superpotential together with $\Phi$.
In general, gauge boson
exchange is only expected in D--terms.
Graviton exchange is forbidden because of gauge symmetry,
as two of the fields must annihilate
into a singlet to allow graviton propagation, which is not
the case for the terms in Eq.\ (37).
Thus $A_4$ is directly related to the coefficient of the
nonrenormalizable term in the
superpotential.
$$W_4={{g}\over{2\pi}}{1\over4}{I\over{M}}(
d_{L_2}^cQ_2h'_{45}{\bar\Phi}_2+e_{L_2}^cL_2h'_{45}{\bar\Phi}_2)\eqno(47)$$
where the relation $${1\over2}g\sqrt{2\alpha^\prime}=
{\sqrt{8\pi}\over M_{\rm Pl}}={1\over2}{1\over M}\eqno(48)$$ has been used.
These quartic order terms in the superpotential will become
effective mass terms for the bottom quark and tau lepton
provided that ${\bar\Phi}_2$ get a VEV of order $M$.
\bigskip
\ll{\bf 6. Calculation of the effective Yukawa couplings}
\smallskip
The massless spectrum of the free fermionic models contains an ``anomalous''
$U(1)$ symmetry. The ``anomalous'' $U(1)$ generates a
Fayet-Iliopoulos $D$--term at the one--loop level in
string perturbation theory.
The Fayet-Iliopoulos $D$--term
breaks supersymmetry at the Planck scale and destabilizes the string vacuum.
The vacuum is stabilized and supersymmetry is restored by giving VEVs to some
Standard Model singlets in the massless string spectrum.
The allowed VEVs are constrained by requiring that the vacuum is
$F$ and $D$ flat.
The set of constraints on the allowed VEVs is
summarized in the following set of equations:
$$\eqalignno{{D_A}&={\sum _{k}}{Q_k^A}{\vert\chi_k\vert}^2+
{g^2\over{192\pi^2}}{Tr(Q_A)}=0&(49a)\cr
D_j&=\sum_{k}Q_k^j\vert\chi_k\vert^2=0{\hskip .3cm}&(49b)\cr
\l W\r&=\l{{\partial W}\over{\partial{\eta_i}}}\r=0&(49c)\cr}$$
where ${\chi}_k$ are the fields that get a VEV and ${Q_k}^j$ is their charge
under the $U(1)_j$ symmetry. The set $\{\eta_j\}$ is the set of all
chiral superfields.

The solution to the set of Eqs.(49a,49b) must be positive definite, since
${\vert\chi_k\vert}^2\ge0$. However, as the total charge of these singlets
must have $Q_A<0$ to cancel the ``anomalous'' $U(1)$ D--term equation, in
many models a phenomenologically realistic solution
was not found [\SLM,\price]. Among
the free fermionic standard--like model, that use the NAHE set to obtain three
generations, the only models that were found to admit a solution are models
with $\Delta_{1,2,3}=1$. These models therefore have cubic level Yukawa
couplings only for $+{2/3}$ charged quarks. Several examples exist
of models with mass terms for both $+{2/3}$ and $-{1/3}$ charged quarks
and which do not seem to admit a phenomenologically viable solution.
This, of course, may be just a reflection of the limited model
search that has been performed to date.

The order of magnitude of the VEVs $\l\chi_j\r$ is determined
by the Fayet--Iliopoulos term. Because the Fayet--Iliopoulos term is
generated at the one--loop level in string perturbation theory,
these VEVs can be naturally suppressed
relative to the string--related scale,
$M\equiv{M_{\rm Pl}}/2\sqrt{8\pi}\approx1.2\cdot10^{18}$ GeV.
The exact suppression factors depend on the details of specific
solutions to the set of $F$ and $D$ flatness constraints.
Consequently, some of the nonrenormalizable, order--$n$ terms
become effective renormalizable terms
with effective Yukawa couplings,
$$\lambda=c_{_n}\left({\langle\phi\rangle\over{M}}\right)^{n-3}.\eqno(50)$$

{}From Eq.\ (37) we observe that in order to obtain nonvanishing bottom
quark and tau lepton mass terms in this specific toy model, we need
to find a solution to the set of $D$ and $F$ constraints with,
$\Phi_1\ne0$ or ${\bar\Phi}_2\ne0$. One explicit solution
to the set of constraints is given by the set
$\{\Phi_{45},{\bar\Phi}_{13},{\Phi}_{13},{\bar\Phi}_2\}$, with
$$\vert\langle\Phi_{45}\rangle\vert^2=
3\vert\langle{\bar\Phi}_2\rangle\vert^2=3\vert\langle\Delta_{13}\rangle\vert=
{{3g^2}\over{16\pi^2}}{1\over{ 2\alpha^\prime}}=
{{3g^4}\over{16\pi^2}}{1\over{ M^2}}\eqno(51)$$
where
$\Delta_{13}=(\vert{\bar\Phi}_{13}\vert^2-\vert\Phi_{13}\vert^2)$.

With this solution, after inserting the VEV of ${\bar\Phi}_2$
and the coefficients of the quartic order correlators, the
effective bottom quark and tau lepton Yukawa couplings are
given by,
$$\lambda_b=\lambda_\tau={{I}\over{32\pi^2}}g^3\approx
0.25g^3.\eqno(52)$$

The top quark mass prediction,
Eq.\ (1), was obtained by taking
$g\sim{1/{\sqrt2}}$ at the unification scale.
The three Yukawa couplings are then run
to the low energy scale by using the MSSM
one--loop RGEs.
The bottom and top quarks masses are given by
$$m_t(\mu)=\lambda_t(\mu)v_1=\lambda_t(\mu)
{{v_0}\over{\sqrt2}}\sin\beta{\hskip 1cm}m_b(\mu)=\lambda_b(\mu)v_2=
\lambda_b(\mu){{v_0}\over{\sqrt2}}\cos\beta,\eqno(53)$$where
$v_0={{2M_W}/{g_2}}=246GeV$
and $\tan\beta={{v_1}/{v_2}}$. The bottom quark mass,
$m_b(M_Z)$ and the $W$-boson mass, $M_{W}(M_{ W})$,
are used to calculate the two electroweak VEVs,
$v_1$ and $v_2$. Using the relation,
$$m_t\approx\lambda_t(M_Z)\sqrt{{{2M_W^2}\over{g^2_2(M_W)}}-
\left({{m_b(M_Z)}\over{\lambda_b(m_Z)}}\right)^2}\eqno(54)$$
the top quark mass prediction, Eq.\ (1), is obtained.

The solution, Eq.\ (51), is of course not unique.
It is important to examine what is the range
of ${\bar\Phi}_2$ and consequently of $\lambda_b(M_{\rm string})$
and $\lambda_\tau(M_{\rm string})$
and how they affect the low energy prediction of the heavy fermion
masses. A simple modification of the above solution is obtained
by adding the field ${\bar\Phi}_1$ to
$\{\Phi_{45},{\bar\Phi}_2,\Phi_{13},{\bar\Phi}_{13}\}$ that were used
in the above solution. The VEVs of $\Phi_{45}$
and $\Delta_{13}$ remain the same. We now
obtain the equation,
$$\vert\l{\bar\Phi}_2\r\vert^2+\vert\l{\bar\Phi}_1\r\vert^2=
{{g^2}\over{16\pi^2}}{1\over{ 2\alpha^\prime}},\eqno(55)$$
and,
$$\l{\bar\Phi}_2\r=\sqrt{{{g^2}\over{16\pi^2}}{1\over{ 2\alpha^\prime}}
-\vert\l{\bar\Phi}_1\r\vert^2}=
{{g^2}\over{4\pi}}M\sqrt{1-{{16\pi^2}
\over{g^4M^2}}\vert\l{\bar\Phi}_1\r\vert^2}~.\eqno(56)$$
Consequently, with this solution, $\l{\bar\Phi}_2\r$ varies between
$$0\le\l{\bar\Phi}_2\r\le{{g^2}\over{4\pi}}{M}~,\eqno(57)$$
and the bottom quark and tau lepton Yukawa couplings vary
accordingly,
$$0\le\lambda_b(M_{\rm string})=\lambda_\tau(M_{\rm string})\le
{{I}\over{32~\pi^2}}g^3
\sqrt{1-{{16\pi^2}\over{g_4M_2}}\vert\l{\bar\Phi}_1\r\vert^2}~.\eqno(58)$$
A lower bound can only be imposed from the physical bottom
quark and tau lepton masses.

\bigskip
\ll{\bf 7. The effect of intermediate matter thresholds}
\smallskip

In the preceding sections we calculated the heavy generation Yukawa
couplings at the string scale. The next step in the analysis of the
fermion masses is to renormalize the Yukawa couplings from the string
scale to the electroweak scale.
The spectrum of massless states and the Yukawa couplings
are those that appear in the specific superstring derived
standard--like toy model. In the proceeding analysis I will
make some motivated assumptions with regard to the mass scales
of various states that exist in the specific string model which is
being analyzed.

Superstring theory in general and free fermionic models in particular
predict that all gauge couplings are unified at the string unification
scale [\ginsparg], which is numerically of the order of [\vadim]
$$M_{\rm string}\approx g_{\rm string}\times5\times10^{17}
{\rm GeV}~,\eqno(59)$$
where $g_{\rm string}$ is the unified gauge coupling. Assuming that
the particle content below the string scale consist only of the MSSM
particle spectrum, results in disagreement with the values
extracted at LEP for $\alpha_{\rm strong}(M_Z)$ and
$\sin^2\theta_W(M_Z)$.
In Ref. [\DF] it was shown, in a wide range of realistic
free fermionic models, that heavy string threshold corrections,
non-standard hypercharge normalizations [\DFM],
light SUSY thresholds or intermediate
gauge structure, do not resolve the disagreement with
$\alpha_{\rm strong}(M_Z)$ and $\sin^2\theta_W(M_Z)$.
Instead, as was previously suggested [\price,\GCU,\Gaillard],
the problem may be resolved in the free fermionic models
due to the existence of additional color triplets and electroweak doublets
beyond the MSSM. Indeed, additional color triplets and electroweak doublets
in vector--like representations, beyond the MSSM, in general appear in the
massless spectrum of the realistic string models. The number of such states
and their mass scales are highly model dependent.
Mass terms for these extra states may arise from cubic
or higher--order non--renormalizable terms in the superpotential.
In general, the masses of the extra states are suppressed relative
to the string scale because of the suppression of the
non--renormalizable terms relative to the cubic level terms.
Additional mass scales that are suppressed relative to the Planck
scales may arise, for example, by condensation in the hidden sector [\scales].
These additional matter thresholds also affect the evolution of the
Yukawa couplings, and it is therefore necessary to include their effect
in the analysis of the fermion masses.

In the superstring derived standard--like models such additional
color triplets and electroweak doublets are obtained
from exotic sectors that arise from the additional vectors $\alpha$,
$\beta$ and $\gamma$.
For example, the model of Ref. [\GCU] is obtained from the model
of Ref. [\TOP] by the change of a GSO phase,
$$c\left(\matrix{\gamma\cr1\cr}\right)=-1\rightarrow
c\left(\matrix{\gamma\cr1\cr}\right)=+1~.\eqno(60)$$
This GSO phase change preserves the 
spectrum and interactions of the massless states which arise from 
the basis vectors $\{{\bf 1},S,b_1,b_2,b_3,\alpha,\beta\}$. 
The states and charges which are generated by these partial
set of basis vectors under the four dimensional gauge are therefore
identical to those in the model of Ref. [\TOP].
The effect of the phase change in Eq. (60)
is to modify the spectrum from sectors which contain the basis 
vector $\pm\gamma$. Thus, this phase change does not affect
the calculation of the Yukawa couplings in the preceding section.  
Therefore, the heavy generation Yukawa couplings in this model are
still given by Eqs.\ (34,58). 
The effect of the GSO phase change, Eq. (60), is to modify 
the massless states from the sectors 
$b_{1}+b_3+\alpha\pm\gamma$ and $b_2+b_3+\beta\pm\gamma$. 
This model contains in its spectrum
two pairs of $(\overline{3},1)_{1/3}$
color triplets from these sectors,
with one-loop beta-function coefficients,
$$b_{D_1,{\bar D}_1,D_2,{\bar D}_2}={\left(\matrix{{1\over2}\cr
                      0\cr
                     {1\over5}\cr}\right)};\eqno(61)$$
one additional pair of color triplets,
$(\overline{3},1)_{1/6}$, from the sector
$1+\alpha+2\gamma$ with,
$$b_{D_3,{\bar D}_3}={\left(\matrix{{1\over2}\cr
                      0\cr
                     {1\over{20}}\cr}\right)};\eqno(62)$$
and three pairs of $(1,2)_{0}$ doublets with
$$b_{\ell,{\bar\ell}}={\left(\matrix{0\cr
                      {1\over2}\cr
                         0\cr}\right)}.\eqno(63)$$

The one--loop and two--loop $\beta$ function coefficients of the
states from the sectors $b_j$, the Neveu--Schwarz sector, and the sector
$b_1+b_2+\alpha+\beta$ are identical to those of the MSSM representations.
Similarly, for any state with standard $SU(5)$ embedding the $\beta$--function
coefficients are the same as for the $SU(5)$ representations.
The two--loop $\beta$ function coefficients of the exotic matter are
$$  b_{D_3,{\bar D}_3}={\left(\matrix{{1\over9}&0&{4\over{15}}\cr
                                      0&0&0\cr
                     {1\over{30}}&0&{{17}\over3}\cr}\right)}~~~{\rm and}~~~
b_{\ell,{\bar\ell}}={\left(\matrix{0&0&0\cr
                      0&{7\over2}&0\cr
                      0&0&0\cr}\right)}.\eqno(64)$$

This particular combination of representations and hypercharge
assignments opens up a sizable window
in which the low--energy data and string unification can
be reconciled.
The standard--like models predict
$\sin^2\theta_W={3/8}$ at the unification scale due to the
embedding of the weak hypercharge in $SO(10)$.
The $SO(10)$ embedding of the weak
hypercharge in these models
enables string scale gauge coupling unification
to be in agreement with the low energy data.
Of course, there exist a large number of possible scenarios
for the mass scales of the extra states and classification
of all these possibilities is beyond the scope of this paper.
It is found, for example, that if the extra triplets,
$\{D_1,{\bar D}_1,D_2,{\bar D}_2,D_3,{\bar D}_3\}$
all have equal masses in the approximate range
$ 2\times 10^{11} \leq M_3 \leq 7\times 10^{13}$ GeV
with the doublet masses in the corresponding range
$ 9\times 10^{13} \leq M_2 \leq 7\times 10^{14}$ GeV,
then agreement with LEP data can be obtained [\DF].

The analysis proceeds as follows.
The heavy generation Yukawa couplings, $\lambda_t$, $\lambda_b$
and $\lambda_\tau$, are renormalized from the string scale to the
electroweak scale by running the two--loop supersymmetric
RGEs for the gauge and
Yukawa couplings, including the contribution of the extra matter.
The Yukawa couplings at
$M_{\rm string}$ are given by Eqs.\ (34,58) in terms of $g_{\rm string}$.
The top quark Yukawa coupling is given by Eq.\ (34).
The bottom quark and tau lepton Yukawa couplings as
a function of $\l{\bar\Phi}_1\r$ are given
by Eq.\ (58).
The string unification scale, $M_{\rm string}$ is determined by Eq.\ (59).
The unified gauge coupling,
$\alpha_{\rm string}$, is varied in the range $0.03-0.07$.
The gauge coupling heavy string threshold corrections in this toy model
were analyzed in Ref. [\DF].
The two--loop coupled supersymmetric RGEs for the gauge and Yukawa couplings
are then evolved to the extra doublets and triplets
thresholds. The three color triplet pairs and three electroweak doublet
pairs, beyond the MSSM, are assumed to be degenerate at the mass scales
$M_3$ and $M_2$ respectively.
The extra doublet and triplet thresholds are varied in the ranges
$$\eqalignno{
   1\times 10^{13} &\leq M_2 \leq 1\times 10^{16}~{\rm GeV}\cr
   9\times 10^{9}  &\leq M_3 \leq 1\times 10^{12} ~{\rm GeV},&(65)\cr}$$
respectively.
The contribution of each threshold to the $\beta$--function coefficients
is removed in a step approximation. The coupled two--loop RGEs are
then evolved to the approximate top quark mass scale, $m_t\approx175$ GeV.
At this scale the top quark Yukawa coupling and
$\alpha_{\rm strong}(m_t)$ are extracted,
and the contribution of the top quark to the RGEs is removed.
The two--loop supersymmetric RGEs
are then evolved to the $Z$ mass scale and agreement with
the experimental values of $\alpha_{\rm strong}(M_Z)$,
                           $\sin^2\theta_W(M_Z)$, and
                           $\alpha^{-1}_{\rm em}(M_Z)$ is imposed.
In this section all the superpartners are assumed to be degenerate at $M_Z$.
Possible corrections due to the superparticle spectrum will be examined
in the next section.
The gauge couplings of
$SU(3)_{\rm color}\times U(1)_{\rm em}$ are then extrapolated from the
$Z$--boson mass scale to the bottom quark mass scale. The running
bottom quark and tau lepton masses are evolved back from their
physical mass scale to the $Z$ mass scale by using the
three--loop QCD and two--loop QED RGEs [\BBO].
The bottom mass is then used to extract the
running top quark mass, using Eq.\ (54).
The physical top quark mass is given by,
$$m_t(physical)=m_t(m_t)(1+{4\over{3\pi}}\alpha_{\rm strong}(m_t)),\eqno(66)$$
where $m_t(m_t)$ is given by Eq.\ (54).

{}From Eq.\ (37) we observe that in this model
$\lambda_b=\lambda_\tau$ at the string unification scale.
Consequently, an additional prediction for the mass ratio
$$
\lambda_b(M_Z)/\lambda_\tau(M_Z)=m_b(M_Z)/m_\tau(M_Z)\eqno(67)
$$
is obtained.
$\lambda_b$ and $\lambda_\tau$ are extrapolated from the string unification
scale to the $Z$--mass scale using the two--loop RGEs with the intermediate
matter thresholds, as described above. The gauge couplings of
$SU(3)_{\rm color}\times U(1)_{\rm em}$ are then extrapolated to the bottom
quark mass scale. The bottom quark and tau lepton masses
are then extrapolated
to the $Z$ mass scale. We can then compare the predicted ratio
on the left--hand side of Eq.\ (67) with the experimentally extrapolated
ratio on the right--hand side.

\bigskip
\noindent{\it Low--Energy Experimental Inputs}
\bigskip

For the subsequent analysis, the input
parameters are the tree-level string
prediction, Eq.\ (59), and $g_{\rm string}$
is varied as described above.
The mass of the $Z$--boson is [\expalpha]
$$M_Z ~\equiv~ 91.161\pm 0.031~{\rm GeV}~.\eqno(68)$$
The RGE's are run from the string scale to the
$Z$ scale. At the $Z$ scale we obtain predictions
for the gauge parameters $\alpha_{\rm strong}$,
$\sin^2\theta_W$ and $\alpha_{\rm em}$. These
predictions are constrained to be in the experimentally
allowed regions [\expalpha],
$$\eqalignno{
  \alpha_{\rm strong}(M_Z)&=0.12\pm0.01,\cr
  \sin^2\theta_W(M_Z)&=0.232\pm0.001\cr
  \alpha^{-1}_{\rm em}(M_Z)&=127.9\pm0.1~.&(69)\cr}$$

Note that these values are obtained in the
${\overline{MS}}$--renormalization scheme
while the predictions from the supersymmetric
RGE's are obtained in the ${\overline{DR}}$--renormalization
scheme. The predictions are converted to the
${\overline{MS}}$--renormalization scheme by using
the conversion factors,
$${1\over{\alpha_i^{\overline{DR}}}}=
  {1\over{\alpha_i^{\overline{MS}}}}-{C_{A_i}\over{12\pi}}\eqno(70)$$
where the $C_{A_i}$ are the quadratic Casimir
coefficients of the adjoint representations
of the gauge factors: $C_{A_3}=3$, $C_{A_2}=2$, $C_{A_1}=0$.

The running bottom quark and tau lepton masses in the
${\overline{MS}}$--renormalization scheme are [\expalpha]
$$\eqalignno{m_b(m_b)&=4.4\pm0.3 ~~~{\rm GeV}~~~~~~~~~{\rm and}\cr
             m_\tau(m_\tau)&=1777.1^{+0.4}_{-0.5} ~{\rm MeV}~.&(71)\cr}$$
These values are extrapolated from the low energy regime to the
$Z$ mass scale using the three--loop QED and two--loop QCD RGEs.
The conversion from $\overline{MS}$ to $\overline{DR}$
increases $m_b$ by roughly half a percent and has virtually no effect on
$m_\tau$ [\hrs].

\bigskip
\noindent{\it Numerical results}
\bigskip

In this section all the superpartners are assumed to be degenerate at
the $Z$ mass scale. Some possible corrections due to the
splitting in the supersymmetric mass spectrum are examined
in the next section. It should be emphasized that the numerical
analysis is not intended as a complete analysis of the
parameter space.
The purpose of the numerical analysis
is to illustrate how stringy calculations may be confronted
with experimental data and what are the still missing pieces
in trying to improve the predictability of the string derived models.

The heavy generation Yukawa couplings are given by
Eqs.\ (34,58). In Fig.\ (1) the dependence of
$\lambda_b(M_{\rm string})=\lambda_\tau(M_{\rm string})$
on $\l{\bar\Phi}_1\r$ is shown. The lower limit arises
from requiring that ${\rm Im}(\lambda_t(M_Z))=0$.
In all of the figures, agreement with the gauge parameters
at the $Z$--boson mass scale is imposed.
In Fig.\ (2), the predicted physical top quark mass $m_t(m_t)$
is plotted versus the predicted value of $\tan\beta$.
{}From the figure we observe that the predicted top quark mass
varies in the interval $90~{\rm GeV}\le m_t\le 205~{\rm GeV}$.
In Fig.\ (3) the top quark mass is plotted versus $\lambda_t(m_t)$.
Although the top Yukawa coupling is near its fixed point, the predicted
top quark mass varies over a wide range. This is of course expected as
it merely reflects the dependence of the calculated top quark mass
on the bottom quark Yukawa coupling,
which is illustrated in Fig.\ (4). The dependence on $\lambda_b(M_Z)$
in turn is a result of the freedom in the determination
of the bottom Yukawa coupling at the sting scale. In Fig.\ (5)
$m_t$ is plotted versus $\l{\bar\Phi}_2\r/M$ which demonstrated the dependence
of the predicted top quark mass on the nonvanishing VEVs in the
DSW mechanism. Similarly, the predicted value of $\tan\beta$
depends on the initial boundary conditions and on $\lambda_b(M_Z)$, which
is shown in fig (6). Fig (7) shows the dependence of $m_t$ on
$\alpha_{\rm strong}(M_Z)$. Again although there is a slight increase
in the predicted values of $m_t$ as $\alpha_{\rm strong}(M_Z)$ increases,
a strong dependence is not observed.

In Fig.\ (8) the predicted ratio $\lambda_b(M_Z)/\lambda_\tau(M_Z)$
is plotted versus the experimentally observed ratio $m_b(M_Z)/m_\tau(M_Z)$.
The bottom quark mass is varied in the interval
$4.1-4.7$ GeV.
It is seen that qualitatively there is very good agreement between
the predicted ratio and the experimentally observed ratio.
Over some of the parameter space the predicted ratio is somewhat
larger than the experimentally observed mass ratio, and better
agreement is obtained for the larger values of $m_b\approx4.5-4.7$ GeV.
It is very important to note that the additional intermediate matter
thresholds that are needed to obtain agreement with the gauge
parameters are also crucial to maintain the agreement with the
$b/\tau$ mass ratio.
This is due to the dependence of the running bottom mass
on the strong coupling.
As the intermediate matter states prevent
the strong coupling from growing outside its experimental
bound, they prevent the bottom quark mass from becoming too large.
Thus, the intermediate matter thresholds that are required for
string gauge coupling unification to be in agreement with the low energy
data are also required for obtaining the correct mass ratio
${{m_b}/{m_\tau}}$.
Of course, the splitting in the supersymmetric
mass spectrum can modify this picture. This will be investigated in the
next section. In Fig.\ (9) $m_{\rm top}$ is plotted
versus the ratio $\lambda_b(M_Z)/\lambda_\tau(M_Z)$.
It is observed that for the explored region of the
parameter space there is no strong dependence of the
predicted top quark mass on the Yukawa ratio.
This again reflects the fact that the top quark
mass mainly depends on the bottom Yukawa, or alternatively
on $\tan\beta$. It is noted that the intermediate matter
thresholds which are required to maintain the low values of
$\alpha_{\rm strong}$, therefore also prevent the bottom Yukawa from
growing too large and consequently maintain the experimentally
viable ratio of $\lambda_b/\lambda_\tau$.

\bigskip
\ll{\bf 8. SUSY breaking effects}
\smallskip

In the analysis in the previous section it was assumed that all
the superpartners are degenerate at the $Z$--boson mass scale.
However, in general the supersymmetric spectrum is split and
may induce substantial threshold corrections to the calculation
of the fermion masses. Furthermore, In the previous analysis
the bottom quark and $W$--boson masses were used to fix the
two electroweak VEVs $v_1$ and $v_2$.
The fermion masses are then calculated in terms of their Yukawa couplings
to the Higgs doublets and in terms of the VEVs of the neutral
component of the electroweak doublets. However, in local supersymmetric
theories, for given boundary conditions at the unification scale ,
the electroweak VEVs can be determined from the
running of the Renormalization Group Equation and minimization of
the one--loop effective potential. In this section,
I briefly examine the minimization of the Higgs potential
and the effects of the heavy superpartner thresholds.
In principle it may be possible to extract the soft SUSY breaking parameters
from the superstring models.
It should be emphasized however that in this paper the derivation
of the SUSY breaking parameters from the superstring models is not
attempted.
The purpose of this section is to briefly examine the potential effects
of the SUSY breaking sector on the fermion mass predictions.
An attempt to extract further information on the SUSY breaking
sector from the string models will be reported in future work.

The analysis proceeds as follows.
The string unification scale is given by Eq.\ (59), and I take
$g_{\rm string}=0.824$.
The spectrum at the string unification scale consist of the MSSM states
plus the additional color triplets and electroweak doublets.
The Renormalization Group Equations are those of the MSSM including the
contribution of the extra matter states.
The parameters of the SUSY
breaking sector consist of the universal trilinear coefficient $A_0$, and
the universal scalar and gaugino masses, $m_0$ and $m_{1/2}$, respectively.
The boundary conditions for the soft SUSY breaking terms
at the unification scale are taken to be universal and are varied over a
sample of the parameter space (see table 3).
\input tables.tex
\vskip .75cm
{\hfill
{\begintable
\ Parameter $X$ \ \|\  $X_i$  \ \|\ $X_f$ \ \|\ $\Delta X$  \crthick
        $A_0$~(GeV)             \|      -200    \|      200     \|   100  \nr
        $m_0$~(GeV)             \| ~~~0         \|      200    \|    100  \nr
        $m_{1/2}$~(GeV)         \| ~100         \|      300    \|    100
\endtable}}
\smallskip
\parindent=0pt
{{\it Table 3 The range and sampling size of the parameter space.
Each free parameter $X$ is sampled in the interval $(X_i,X_f)$ with
spacing $\Delta X$ between consecutive points.}
\vskip 1cm
\parindent=15pt

The heavy generation Yukawa couplings at the unification scale
are given by Eqs.\ (34,58).
Similar to the procedure described in the previous section,
the RGEs are evolved from the string unification scale to the electroweak
scale. The contribution of the extra color triplets and electroweak
doublets to the $\beta$--function coefficients is removed in a step
approximation. As a specific example, the three additional color
triplet pairs are taken to be degenerate at
$ M_3 = 2.8\times 10^{11}$ GeV
with the three pairs of electroweak doublets degenerate at
$ M_2 = 4.0\times 10^{13}$ GeV.
Agreement with the low energy observables
$\alpha_{\rm em}(M_Z)$, ${\sin\theta_W}(M_Z)$ and $\alpha_s(M_Z)$ is
then obtained.

The analysis of electroweak symmetry breaking and
the minimization of the Higgs scalar potential is
standard and has been examined extensively in the
context of the MSSM [\mssmewx].
Below the intermediate scales of the additional vector--like states
the matter spectrum is that of the MSSM. The Higgs part of the
MSSM scalar potential is given by,
$$\eqalignno{V(H_1,H_2)=&(m_{H_1}^2+\mu^2)\vert{H_1}\vert^2+
             (m_{H_2}^2+\mu^2)\vert{H_2}\vert^2+
             B\mu(H_1H_2+h.c.)+\cr
   &{1\over8}g_2^2({H_1^\dagger}\sigma{H_1}+{H_2^\dagger}\sigma{H_2})^2+
   {1\over8}g^{\prime^2}(\vert{H_2}\vert^2-\vert{H_2}\vert^2)^2+
        \Delta V,&(72)}$$
where
$H_1\equiv{{H^0_1\choose H^-_1}}$ and
$H_2\equiv{{H^+_2\choose H^0_2}}$ are the two complex Higgs fields,
$m_{H_1}^2$, $m_{H_2}^2$ and $B$ ($\mu$) are the soft supersymmetric
mass parameters renormalized down to the weak scale,
$m_3\equiv B\mu<0$ and $g_2$ and $g^\prime$ are the $SU(2)_L$ and
$U(1)_Y$ gauge couplings respectively. The one--loop correction
$\Delta V$ is given by
$${\Delta V={1\over64\pi^2}{\rm STr}\,{\cal M}^4
\left(\ln{{\cal M}^2\over Q^2}-{3\over2}\right),}\eqno(73)$$
where ${\rm STr}\,f({\cal M}^2)
=\sum_j(-1)^{2j}(2j+1){\rm Tr}\,f({\cal M}^2_j)$ and ${\cal M}^2_j$ are the
field--dependent spin-$j$ mass matrices. The one--loop corrections, $\Delta V$,
receive contributions from all particle species. It is sufficient however
to include only the corrections due to the heavy particles, i.e. the heavy
top quark and the heavy superpartners.
Requiring a negative eigenvalue to the neutral Higgs mass squared matrix
and that the Higgs potential is bounded from below imposes two
conditions on the running of the mass parameters:
The well known tree level minimization constraints
$$\eqalignno{&1.~~m_1^2m_2^2-m_3^4<0\cr
  &2.~~m_1^2+m_2^2-2\vert{m_3^2}\vert>0,\cr}$$
where $m_{1,2}^2=m_{h_{1,2}}^2+\mu^2$ and $m_3^2=B\mu$.
Similarly, the constraint that the vacuum
does not break color and electric charges is imposed [\cm].

Ordinarily, at this stage, to minimize the Higgs potential one
would fix the Higgs mixing parameter $\mu$ and the bilinear
coupling $B$ from their initial values at the unification scale and
their scaling to the weak scale. The electroweak VEVs, $v_1$ and $v_2$,
can then be obtained from minimization
of the one--loop effective potential. Here however a different procedure
is followed. The VEVs $v_1$ and $v_2$ are fixed by using the physical
bottom quark and $W$--boson masses at the $Z$ scale.
The one--loop potential is then minimized for the $\mu$ and $B$
parameters. An initial guess for the minimization values of
$\mu$ and $B$ is obtained from the two minimization conditions
of the tree level superpotential, Eq.\ (72).
The sparticle spectrum is obtained by using the
regular parameterization (see for example Ref. [\FG] for the notation
used in this paper).
In the analysis of the sparticle spectrum the Yukawa couplings
of the two light generations are neglected. The heavy generation mass
eigenstates are obtained by diagonalizing the respective $2\times 2$
mass matrices. Similarly the neutralino, chargino and Higgs mass eigenstates
are obtained by diagonalizing the respective
$4{\times}4$ and $2{\times}2$ mass
matrices.
The numerical contributions to the tree--level and one--loop Higgs potential
are obtained from Eqs.\ (72,73).
The one--loop effective Higgs potential is then
minimized numerically by varying the $\mu$ and $B$ parameters.
In this procedure $\mu$ and $B$ become the computed parameters at the
weak scale. This is possible because the running of the SUSY parameters
does not depend on the values $\mu$ and $B$ (except for $\mu$ itself).
The supersymmetric mass spectrum is then recomputed using the minimizing
values of $\mu$ and $B$.

The above analysis demonstrates that in principle the predicted
values of the top quark mass and the mass ratio $m_b(M_Z)/m_\tau(M_Z)$
can be compatible with the minimization of the one--loop effective
Higgs potential. Whether this is indeed realized in the string models
awaits further analysis of the SUSY breaking sectors in these models
and of the $\mu$ parameter.
The $\mu$ problem is one of the more challenging problems facing
supersymmetry and superstring phenomenology. Naive solutions
to this problem can be contemplated both in field theoretic supersymmetry
and in superstring theory [\musolutions]. For example, in the superstring
derived standard--like models a possible solution was proposed in which
the $\mu$ term is generated from a nonrenormalizable term in the
superpotential [\NRT]. The nonrenormalizable term contains VEVs that break the
$U(1)_{Z^\prime}$, of the $SO(10)$ group, which is orthogonal to the
Standard Model weak hypercharge. It is argued that if the $U(1)_{Z^\prime}$
symmetry is broken at an intermediate energy scale then a $\mu$ parameter
of the required order of magnitude can be generated. Additional symmetries
that arise from the compactified degrees of freedom prevent any other
$\mu$ term from being generated. However, in general, in a realistic
solution of the string vacuum most of these additional symmetries
need to be broken in order to generate potentially realistic
fermion mass matrices [\CKM]. Thus, in a generic realistic solution there
is the danger that a $\mu$ term of the order of
$({\l\Phi\r}^{n-1}/M^n)$ will be generated from an order $n$
nonrenormalizable term. It is not unconceivable however that in
some string models a residual discrete symmetry
or a remnant of a custodial symmetry [\custodial] will remain unbroken and
will prevent a large $\mu$ term from being generated. Such
symmetries in a specific string model have been shown for
example to prevent proton decay from dimension four operators
to all orders of nonrenormalizable terms. However,
whether such a scenario can be realized, is not only model
dependent but also depends on the details of the vacuum shift
in the application of the DSW vacuum shift.

The splitting in the supersymmetric mass spectrum
also induces corrections to the tree level mass predictions.
The supersymmetric thresholds affects both the gauge coupling
and the Yukawa couplings. The threshold corrections
for the gauge couplings and Yukawa couplings received
considerable attention in the context of the MSSM [\yukthresh].
It is not the purpose here to present a detailed numerical
investigation of the supersymmetric threshold effects
and in general the corrections are expected to be small.
I examine the correction of the bottom quark mass
due to the gluino and Higgsino. These corrections have been
shown to be important in the case of large $\tan\beta$ [\hrs,\largetanbeta].
The corrected bottom quark mass is given by
$$m_b=\lambda_bv_1(1+\Delta(m_b))\eqno(74)$$
where $\Delta(m_b)$ receives contributions coming
from bottom squark--gluino loops and top squark--chargino
loops, and is given by
$$\eqalignno{
\Delta(m_b)&={{2\alpha_3}\over{3\pi}}m_{\tilde g}\mu\tan\beta
I(m_{{\tilde b},+}^2,m_{{\tilde b},-}^2,m_{\tilde g}^2)\cr
&+{\lambda_t\over{4\pi}}A_t\mu\tan\beta
I(m_{{\tilde t},+}^2,m_{{\tilde t},-}^2,\mu^2),&(75)\cr}$$
where $m_{\tilde g}$ is the gluino mass, $m_{{\tilde b},\pm}$ and
$m_{{\tilde t},\pm}$ are the sbottom and stop mass eigenstates
respectively, and $A_t$ is the .
The integral function is given by,
$$I(a,b,c)={{ab\ln(a/b)+bc\ln(b/c)+ac\ln(c/a)}\over
{(a-b)(b-c)(a-c)}}~.\eqno(76)$$

In Fig.\ (10) the corrected bottom quark--tau lepton mass ratio
is plotted versus the experimentally extrapolated mass ratio
$m_b(M_Z)/m_\tau(M_Z)$ for the sample of points in the parameter
space which are given in table 3. In Fig.\ (11) the predicted mass
ratio at the $Z$ scale, divided by the experimentally extrapolated
mass ratio, is plotted versus the experimentally extrapolated mass
ratio. The bottom quark mass is varied in the range
$4.1-4.7$~GeV. The predicted mass ratio is seen to vary
between $\approx1-1.2$. Thus, the predicted mass ratio is
in reasonable agreement with the experimental mass ratio and
better agreement is obtained for the larger values of $m_b(m_b)$.

\bigskip
\ll{\bf 9. Discussion and conclusions}

The nature of the electroweak symmetry breaking mechanism and
the origin of the fermion masses is one of the important pieces in
the puzzle of elementary particle physics.
The calculation of the fermion masses from fundamental
principles is therefore an important task.
Within the context of theories of unification,
the fermion masses are expected to arise due to
some underlying Planck scale physics.
Superstring theory provides at present
the best tool to probe Planck scale physics.

The superstring derived standard--like models in the free fermionic
formulation possess many attractive properties. An important property
of the these models is the possible solution to the problem
of proton stability [\PD]. A second important property of free
fermionic models is the prediction
$\sin^2\theta_W={3/8}$ at the unification scale due to the
embedding of the weak hypercharge in $SO(10)$.
This rather common result from the point of view of
regular GUTs is highly nontrivial from the point of view
of string models. It is only due to the standard
$SO(10)$ embedding of the weak hypercharge that
free fermionic models can be in agreement with the low
energy data. Recently, it was suggested that
$\sin^2\theta_W(M_U)={3/8}$ is the preferred value also from
considerations of the fermion mass spectrum [\ramond].

Another important property of the superstring derived standard--like models
is the existence of three and only three chiral generation in the massless
spectrum. This property enhances the predictability of the superstring
standard--like models. As a result it is possible to identify the states
in the string models with the physical mass eigenstates of the Standard Model.

In this paper I discussed in detail the calculation of the
heavy generation masses in the superstring derived standard--like models.
In these models the top quark gets a cubic level mass term
while the mass terms for the lighter quarks and leptons are
obtained from nonrenormalizable terms.
The top quark Yukawa coupling and the quartic order correlator of the
bottom quark and tau lepton mass terms were calculated in a specific
model. The numerical coefficient of the quartic order correlator was
calculated explicitly and was shown to be nonzero.
The quartic order mass terms produce effective Yukawa couplings
for the bottom quark and tau lepton after application of the
DSW mechanism. The dependence of the effective Yukawa couplings
on the VEV in the DSW mechanism and the implication on the top
quark mass prediction was studied in detail.
The string--scale coupling unification requires the existence
of intermediate matter thresholds. The gauge
and Yukawa couplings were run from the string unification
scale to the low energy scale in the presence of the intermediate matter.
It was shown that LEP precision data for $\alpha_{\rm strong}$,
$\sin^2\theta_W$ and $\alpha_{\rm em}$ as well as
the CDF/D0 top quark observation and the $b/\tau$ mass relation
can all simultaneously be compatible with the superstring derived
standard--like models.

Although the
calculations were presented in a specific toy model, the features
of this toy model that are relevant for the analysis are shared
by a large class of superstring standard--like models in the
free fermionic formulation. Thus, the results are expected to
hold in the larger class of models. Similar analysis
can of course also be carried out in other semi--realistic string
models.
The above results  motivate further analysis of the fermion masses
in the superstring derived standard--like models. Several
directions should be pursued. The first is to try to extend
the analysis to the lighter generations. Potential
charm quark mass terms appear at the quintic order of the
superpotential. For example,
$u_2Q_2{\bar h^\prime}_{45}{\bar\Phi}_{23}{\Phi_{45}}$,
can provide a quintic order charm quark mass term. The
analysis for such a term involves higher order Ising model
correlators and a two dimensional complex integration.
Another important direction is the analysis of the
supersymmetry breaking sector in the standard--like
models and possible corrections due to the dilaton and
moduli dependence of the Yukawa couplings [\LNZ]. Additional
corrections may arise from the infinite tower of heavy string modes [\stcy].
Thus, although much more work is needed to understand how
the specific string parameters are fixed by the string physics,
we have made the initial steps toward the quantitative
confrontation of string models with experimental data.

\bigskip
\ll{\bf Acknowledgments}
It is a pleasure to thank Keith Dienes, Jogesh Pati and Pierre Ramond
for very stimulating and enjoyable discussions. I would like
to thank the Institute for Advanced Study and
the Institute for Theoretical Physics at Santa Barbara for their
support and hospitality during the initial stages of this work.
This work is supported in part by the Department
of Energy under contract DE--FG05--86ER--40272.
\smallskip
\vfill
\eject
\baselineskip=12pt
\refout
\baselineskip=16pt
\vfill
\eject

\input tables.tex
\nopagenumbers

\magnification=1000
\tolerance=1200

{\hfill
{\begintable
\  \ \|\ ${\psi^\mu}$ \ \|\ $\{{\chi^{12};\chi^{34};\chi^{56}}\}$  \ \|\
{${\bar\psi}^1$, ${\bar\psi}^2$, ${\bar\psi}^3$,
${\bar\psi}^4$, ${\bar\psi}^5$, ${\bar\eta}^1$,
${\bar\eta}^2$, ${\bar\eta}^3$} \ \|\
{${\bar\phi}^1$, ${\bar\phi}^2$, ${\bar\phi}^3$, ${\bar\phi}^4$,
${\bar\phi}^5$, ${\bar\phi}^6$, ${\bar\phi}^7$, ${\bar\phi}^8$} \crthick
${\alpha}$
\|\ 0 \|
$\{0,~0,~0\}$  \|
1, ~~1, ~~1, ~~0, ~~0, ~~0, ~~0, ~~0 \|
1, ~~1, ~~1, ~~1, ~~0, ~~0, ~~0, ~~0 \nr
${\beta}$
\|\ 0 \| $\{0,~0,~0\}$  \|
1, ~~1, ~~1, ~~0, ~~0, ~~0, ~~0, ~~0 \|
1, ~~1, ~~1, ~~1, ~~0, ~~0, ~~0, ~~0 \nr
${\gamma}$
\|\ 0 \|
$\{0,~0,~0\}$  \|
 ~~$1\over2$, ~~$1\over2$, ~~$1\over2$, ~~$1\over2$,
{}~~$1\over2$, ~~$1\over2$, ~~$1\over2$, ~~$1\over2$ \| $1\over2$, ~~0, ~~1,
{}~~1,
{}~~$1\over2$,
{}~~$1\over2$, ~~$1\over2$, ~~0 \endtable}
\hfill}
\smallskip
{\hfill
{\begintable
\  \ \|\
${y^3y^6}$,  ${y^4{\bar y}^4}$, ${y^5{\bar y}^5}$,
${{\bar y}^3{\bar y}^6}$
\ \|\ ${y^1\omega^6}$,  ${y^2{\bar y}^2}$,
${\omega^5{\bar\omega}^5}$,
${{\bar y}^1{\bar\omega}^6}$
\ \|\ ${\omega^1{\omega}^3}$,  ${\omega^2{\bar\omega}^2}$,
${\omega^4{\bar\omega}^4}$,  ${{\bar\omega}^1{\bar\omega}^3}$ \crthick
${\alpha}$ \|
1, ~~~1, ~~~~1, ~~~~0 \|
1, ~~~1, ~~~~1, ~~~~0 \|
1, ~~~1, ~~~~1, ~~~~0 \nr
${\beta}$ \|
0, ~~~1, ~~~~0, ~~~~1 \|
0, ~~~1, ~~~~0, ~~~~1 \|
1, ~~~0, ~~~~0, ~~~~0 \nr
${\gamma}$ \|
0, ~~~0, ~~~~1, ~~~~1 \|\
1, ~~~0, ~~~~0, ~~~~0 \|
0, ~~~1, ~~~~0, ~~~~1 \endtable}
\hfill}
\smallskip
\parindent=0pt
\hangindent=39pt\hangafter=1
\baselineskip=18pt

{{\it Table 1.} A three generations
${SU(3)\times SU(2)\times U(1)^2}$ model.
The choice of generalized GSO coefficients is:
$${c\left(\matrix{b_j\cr
                                    \alpha,\beta,\gamma\cr}\right)=
 -c\left(\matrix{\alpha\cr
                                    1\cr}\right)=
 -c\left(\matrix{\alpha\cr
                                    \beta\cr}\right)=
 -c\left(\matrix{\beta\cr
                                    1\cr}\right)=
  c\left(\matrix{\gamma\cr
                                    1\cr}\right)=
 -c\left(\matrix{\gamma\cr
                                   \alpha,\beta\cr}\right)=
-1}$$ (j=1,2,3), with the others specified by modular invariance
and space--time supersymmetry. $\Delta_{1,2,3}=1\Rightarrow$ cubic
level Yukawa couplings are
obtained only for ${+{2\over3}}$ charged quarks. }
\vskip 2cm

\vfill
\eject

\magnification=1000
\tolerance=1200

{\hfill
{\begintable
\  \ \|\ ${\psi^\mu}$ \ \|\ $\{{\chi^{12};\chi^{34};\chi^{56}}\}$ \ \|\
{${\bar\psi}^1$, ${\bar\psi}^2$, ${\bar\psi}^3$,
${\bar\psi}^4$, ${\bar\psi}^5$, ${\bar\eta}^1$,
${\bar\eta}^2$, ${\bar\eta}^3$} \ \|\
{${\bar\phi}^1$, ${\bar\phi}^2$, ${\bar\phi}^3$, ${\bar\phi}^4$,
${\bar\phi}^5$, ${\bar\phi}^6$, ${\bar\phi}^7$, ${\bar\phi}^8$} \crthick
${\alpha}$
\|\ 1 \| $\{1,~0,~0\}$ \|
1, ~~1, ~~1, ~~1, ~~1, ~~1, ~~0, ~~0 \|
0, ~~0, ~~0, ~~0, ~~0, ~~0, ~~0, ~~0 \nr
${\beta}$
\|\ 1 \|
$\{0,~0,~1\}$  \|
1, ~~1, ~~1, ~~0, ~~0, ~~1, ~~1, ~~0 \|
1, ~~1, ~~1, ~~1, ~~0, ~~0, ~~0, ~~0 \nr
${\gamma}$
\|\ 1 \| $\{0,~1,~0\}$  \|
{}~~$1\over2$, ~~$1\over2$, ~~$1\over2$, ~~$1\over2$,
{}~~$1\over2$, ~~$1\over2$, ~~$1\over2$, ~~$1\over2$ \|
 $1\over2$, ~~0, ~~1, ~~1,
{}~~$1\over2$,
{}~~$1\over2$, ~~$1\over2$, ~~0 \endtable}
\hfill}
\smallskip
{\hfill
{\begintable
\   \ \|\
${y^3y^6}$,  ${y^4{\bar y}^4}$, ${y^5{\bar y}^5}$,
${{\bar y}^3{\bar y}^6}$
\ \|\ ${y^1\omega^6}$,  ${y^2{\bar y}^2}$,
${\omega^5{\bar\omega}^5}$,
${{\bar y}^1{\bar\omega}^6}$
\ \|\ ${\omega^1{\omega}^3}$,  ${\omega^2{\bar\omega}^2}$,
${\omega^4{\bar\omega}^4}$,  ${{\bar\omega}^1{\bar\omega}^3}$ \crthick
${\alpha}$ \|
1, ~~~0, ~~~~0, ~~~~1 \|
0, ~~~0, ~~~~1, ~~~~0 \|
0, ~~~0, ~~~~1, ~~~~0 \nr
${\beta}$\|
0, ~~~0, ~~~~0, ~~~~1 \|
0, ~~~1, ~~~~0, ~~~~1 \|
1, ~~~0, ~~~~1, ~~~~0 \nr
${\gamma}$\|
0, ~~~0, ~~~~1, ~~~~1 \|\
1, ~~~0, ~~~~0, ~~~~1 \|
0, ~~~1, ~~~~0, ~~~~0 \endtable}
\hfill}
\smallskip

\parindent=0pt
\hangindent=39pt\hangafter=1
\baselineskip=18pt

{{\it Table 2.} A three generations
${SU(3)\times SU(2)\times U(1)^2}$ model.
The choice of generalized GSO coefficients is:
$${c\left(\matrix{{\alpha}\cr
                                    b_j,\beta\cr}\right)=
 -c\left(\matrix{{\alpha}\cr
                                    1\cr}\right)=
  c\left(\matrix{\beta\cr
                                    1\cr}\right)=
  c\left(\matrix{\beta\cr
                                    b_j\cr}\right)=
 -c\left(\matrix{\beta\cr
                                    \gamma\cr}\right)=
  c\left(\matrix{\gamma\cr
                                    b_2\cr}\right)=
 -c\left(\matrix{\gamma\cr
                                    b_1,b_3,{\alpha},\gamma\cr}\right)=
-1}$$ (j=1,2,3), with the others
specified by modular invariance and space--time supersymmetry.
Trilevel Yukawa couplings are obtained for ${+{2/3}}$
charged quarks as well as ${-{1/3}}$ charged quarks
and for charged leptons.
$\Delta_1=1\Rightarrow$ Yukawa coupling
for $+{2/3}$ charged quark from the sector $b_1$.
$\Delta_{2,3}=0\Rightarrow$ Yukawa couplings for $-1/3$ charged quarks and
charged leptons from the sectors $b_2$ and $b_3$.}
\vskip 1.5cm

\vfill
\eject

%
\midinsert
\epsfxsize 4.5 truein
\centerline{\epsffile{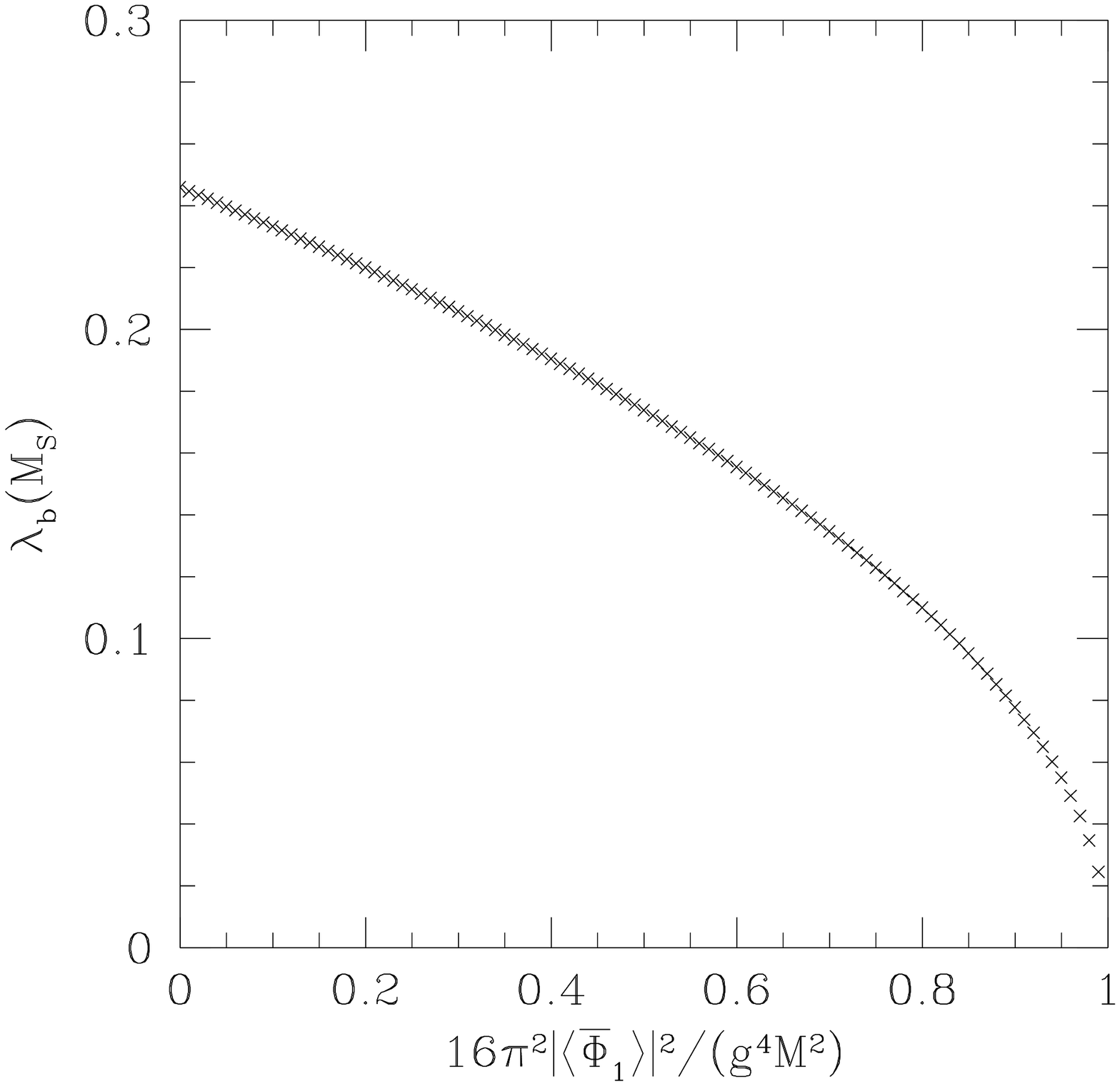}}
\nobreak
\narrower
\singlespace
\noindent
Figure 1. The bottom quark Yukawa coupling at the string unification scale
as a function of the VEV $\l\Phi_1\r$.

\medskip
\endinsert

%
\midinsert
\epsfxsize 4.5 truein
\centerline{\epsffile{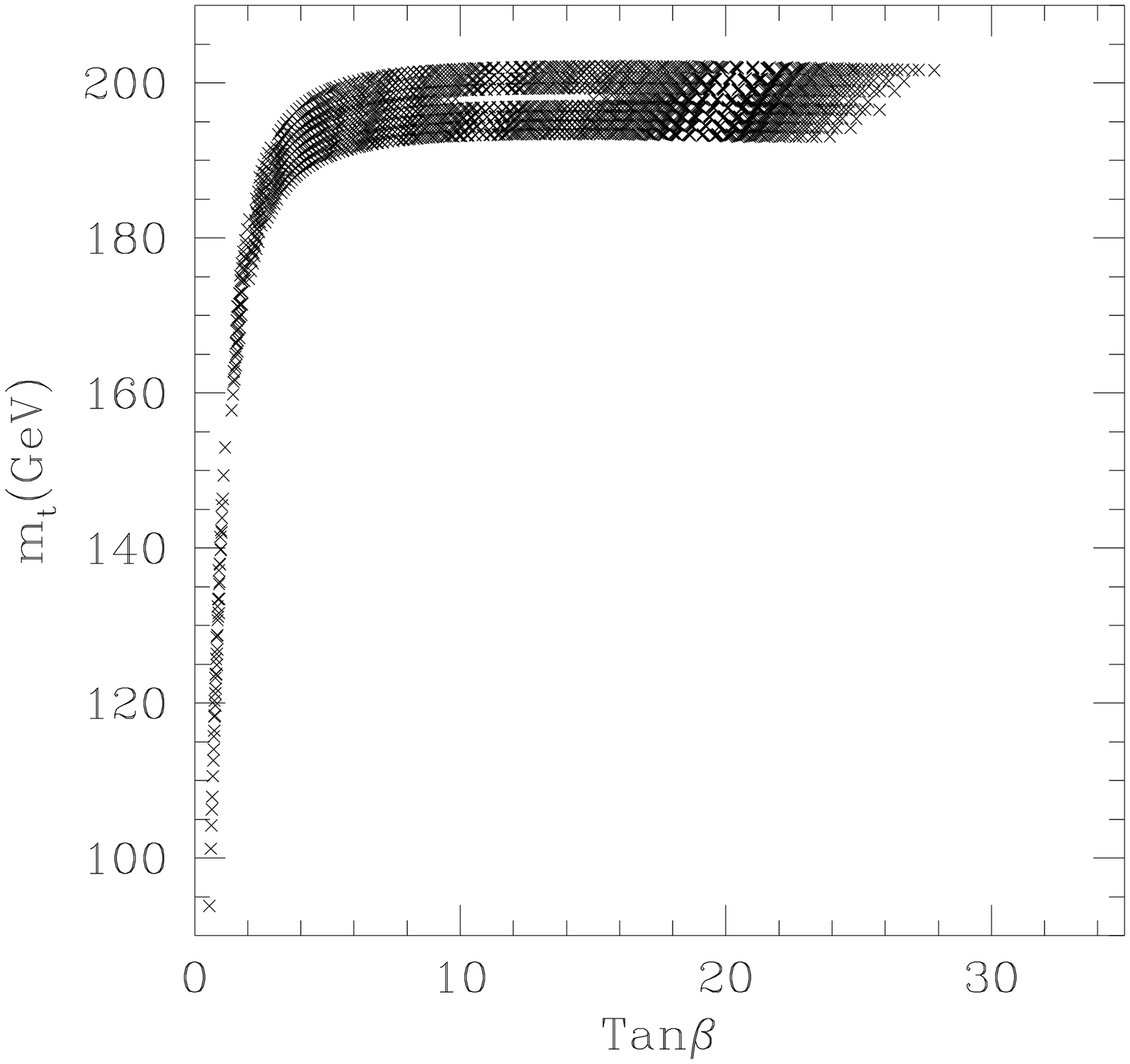}}
\nobreak
\narrower
\singlespace
\noindent
Figure 2. A scatter plot of the physical top quark mass, $m_t(m_t)$
versus the electroweak VEVs ratio, $\tan\beta=v_1/v_2$. Each point
in this plot represents a specific choice for the initial boundary
conditions for the gauge and Yukawa couplings at the string unification
scale and for the mass scales of the intermediate matter states.

\medskip
\endinsert

%
\midinsert
\epsfxsize 4.5 truein
\centerline{\epsffile{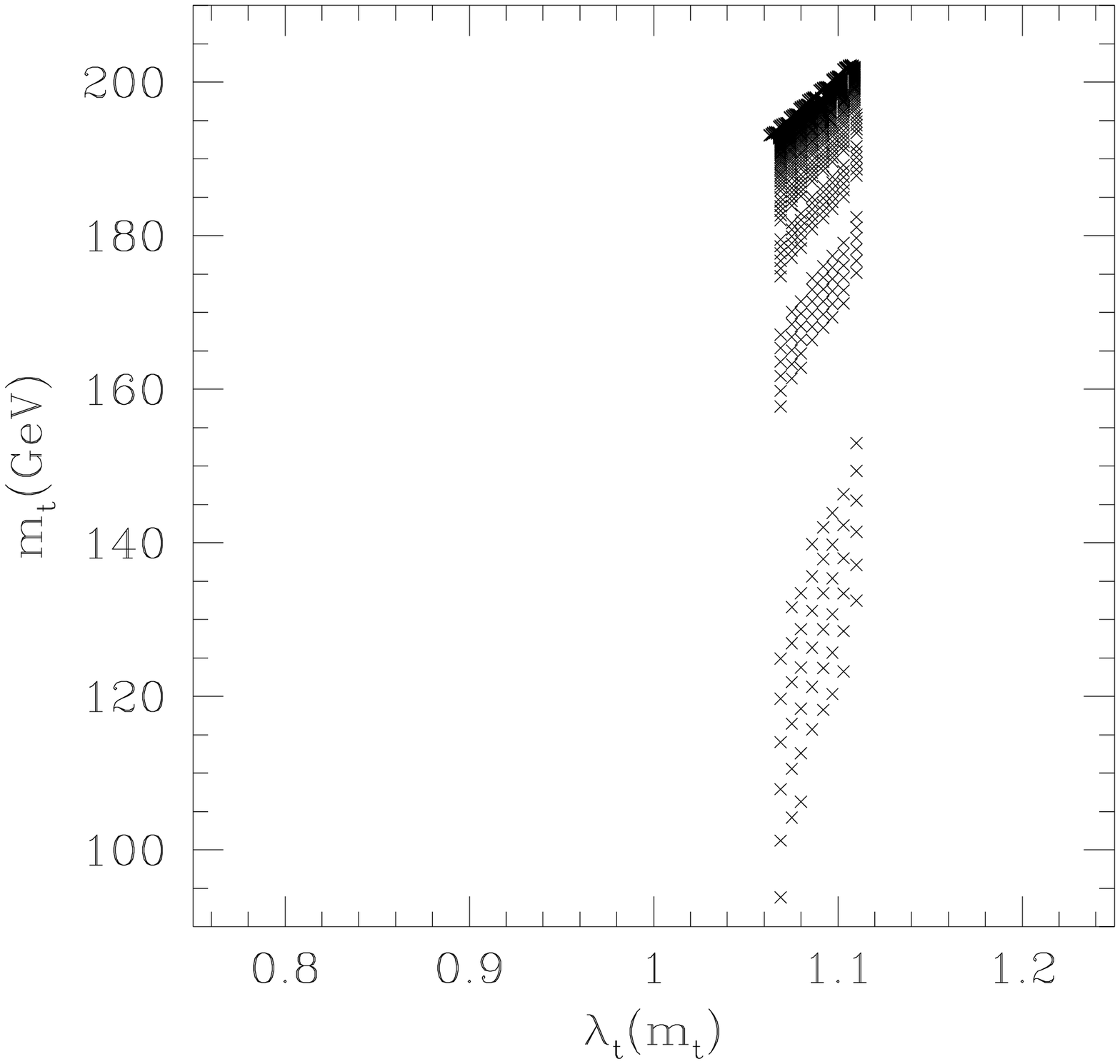}}
\nobreak
\narrower
\singlespace
\noindent
Figure 3. A scatter plot of the physical top quark mass, $m_t(m_t)$
          versus the top quark Yukawa coupling $\lambda_t(m_t)$.
Each point represents a choice of parameters as in figure 2.
$\lambda_t(m_t)$ is found near its fixed point. However, there is
a wide variation in $m_t(m_t)$. This reflects the dependence
of the predicted top quark mass on the bottom quark Yukawa
coupling, or alternatively on the VEVs ratio $\tan\beta$.

\medskip
\endinsert

%
\midinsert
\epsfxsize 4.5 truein
\centerline{\epsffile{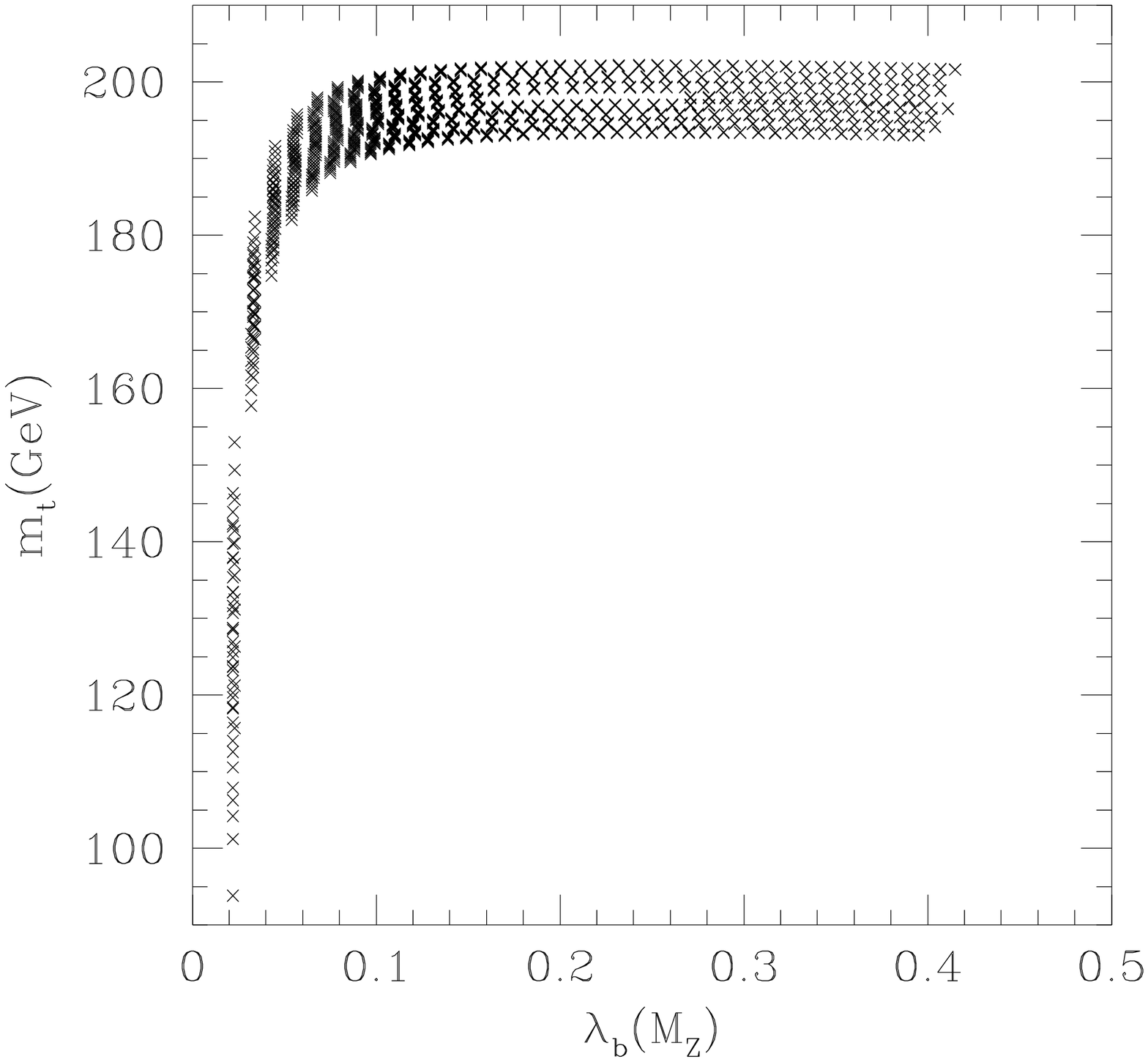}}
\nobreak
\narrower
\singlespace
\noindent
Figure 4. A scatter plot of the physical top quark mass, $m_t(m_t)$
          versus the bottom quark Yukawa coupling $\lambda_b(M_Z)$.
          Each point represents a choice of parameters as in figure 2.
\medskip
\endinsert

%
\midinsert
\epsfxsize 4.5 truein
\centerline{\epsffile{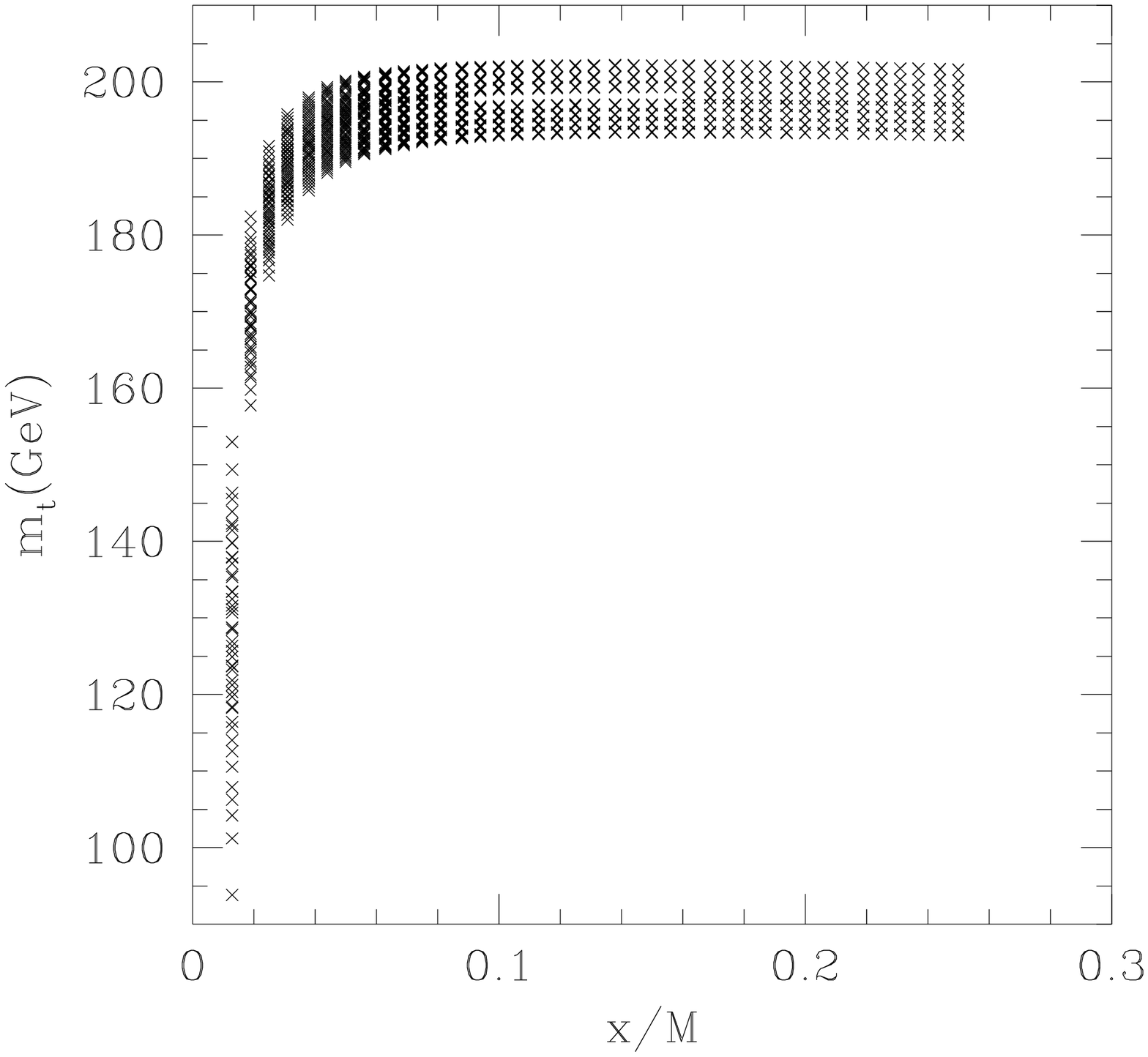}}
\nobreak
\narrower
\singlespace
\noindent
Figure 5. A scatter plot of the physical top quark mass, $m_t(m_t)$
          versus the VEV in the DSW mechanism, $\l{\bar\Phi}_2\r$,
          which fixes the effective bottom quark and tau lepton Yukawa
          couplings at the unification scale.
          Each point represents a choice of parameters as in figure 2.
\medskip
\endinsert

%
\midinsert
\epsfxsize 4.5 truein
\centerline{\epsffile{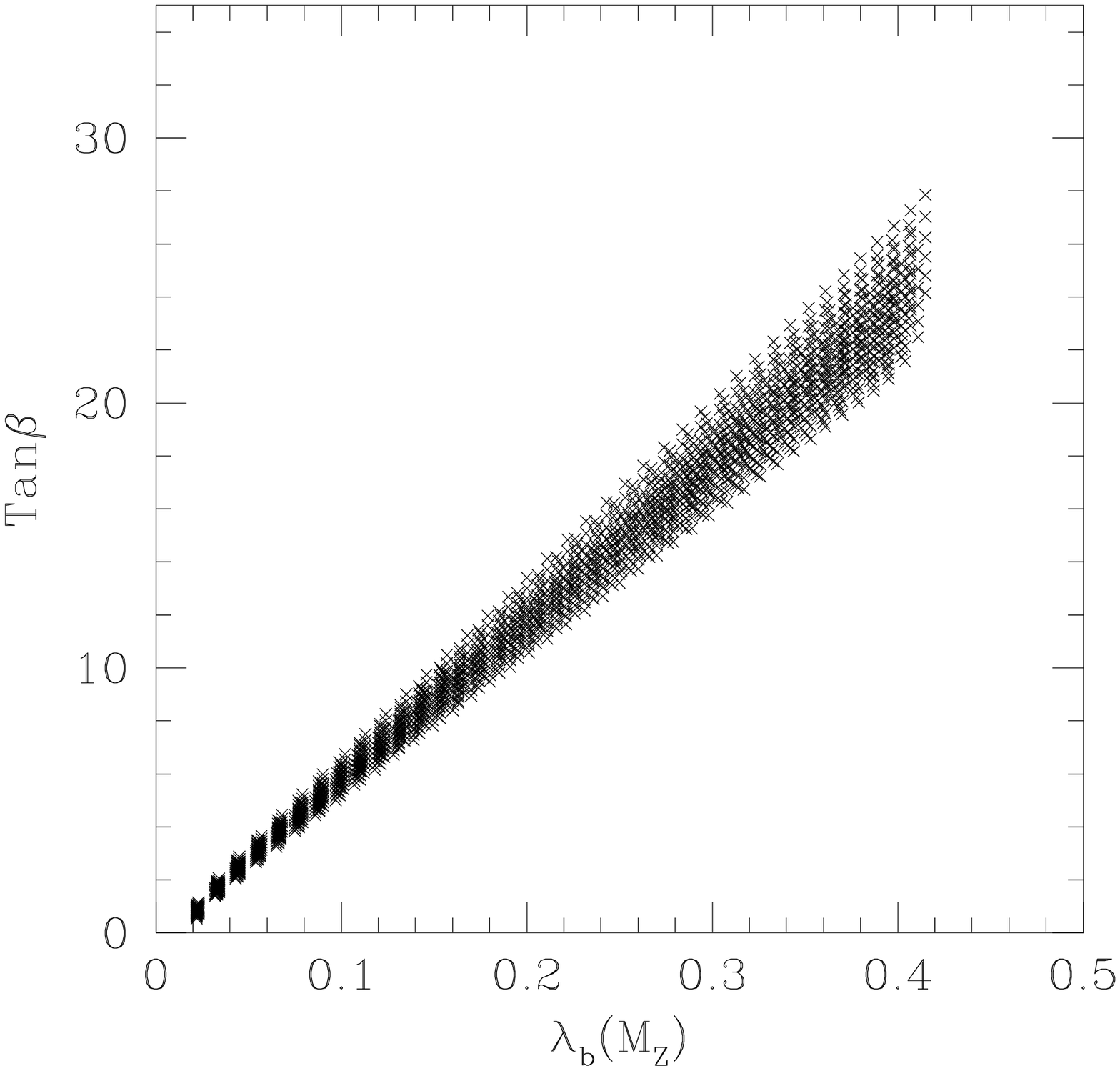}}
\nobreak
\narrower
\singlespace
\noindent
Figure 6. A scatter plot of $\tan\beta$ versus $\lambda_b(M_Z)$.
Each point represents a choice of parameters as in figure 2.
\medskip
\endinsert

%
\midinsert
\epsfxsize 4.5 truein
\centerline{\epsffile{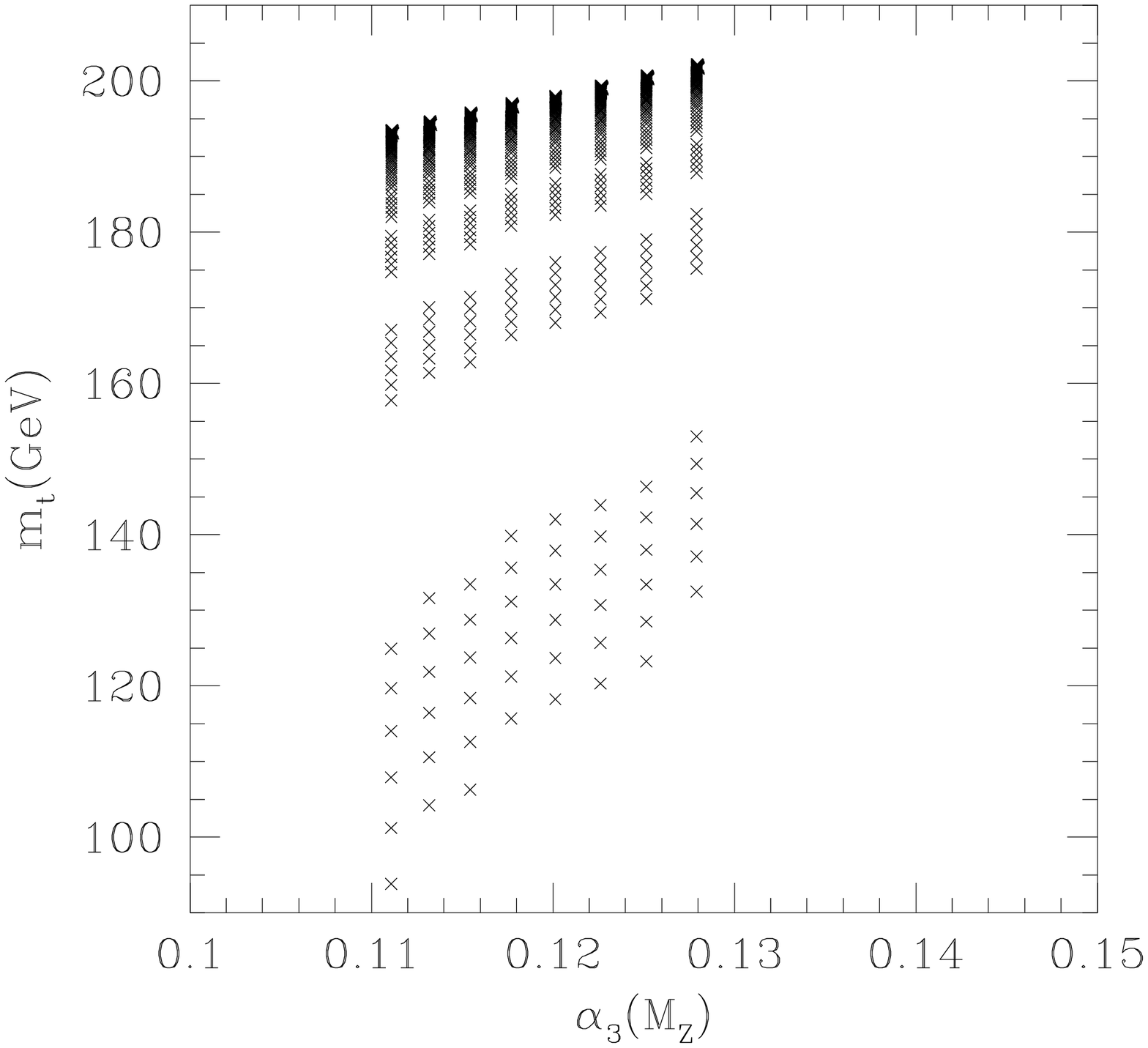}}
\nobreak
\narrower
\singlespace
\noindent
Figure 7. A scatter plot of the physical top quark mass, $m_t(m_t)$
          versus the strong coupling $\alpha_{\rm strong}(M_Z)$.
          Each point represents a choice of parameters as in figure 2.
\medskip
\endinsert

%
\midinsert
\epsfxsize 4.5 truein
\centerline{\epsffile{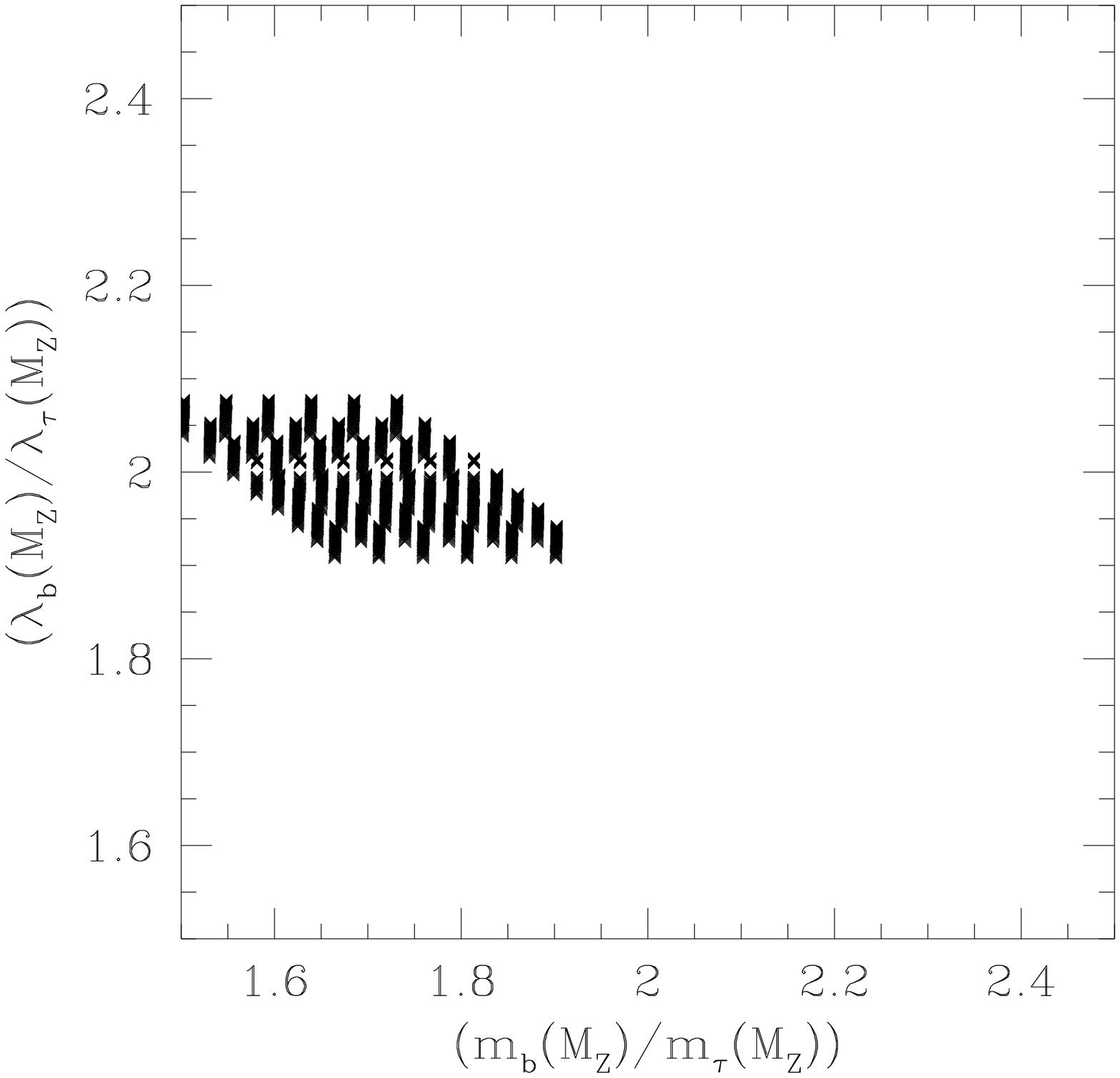}}
\nobreak
\narrower
\singlespace
\noindent
Figure 8. A scatter plot of the predicted ratio
$({\lambda_b(M_Z)/\lambda_\tau(M_Z)})$ versus the experimentally
extrapolated mass ratio $(m_b(M_Z)/m_\tau(M_Z))$. All the superpartners
are assumed to be degenerate at the $Z$ mass scale.
Each point represents a choice of parameters as in figure 2.

\medskip
\endinsert

%
\midinsert
\epsfxsize 4.5 truein
\centerline{\epsffile{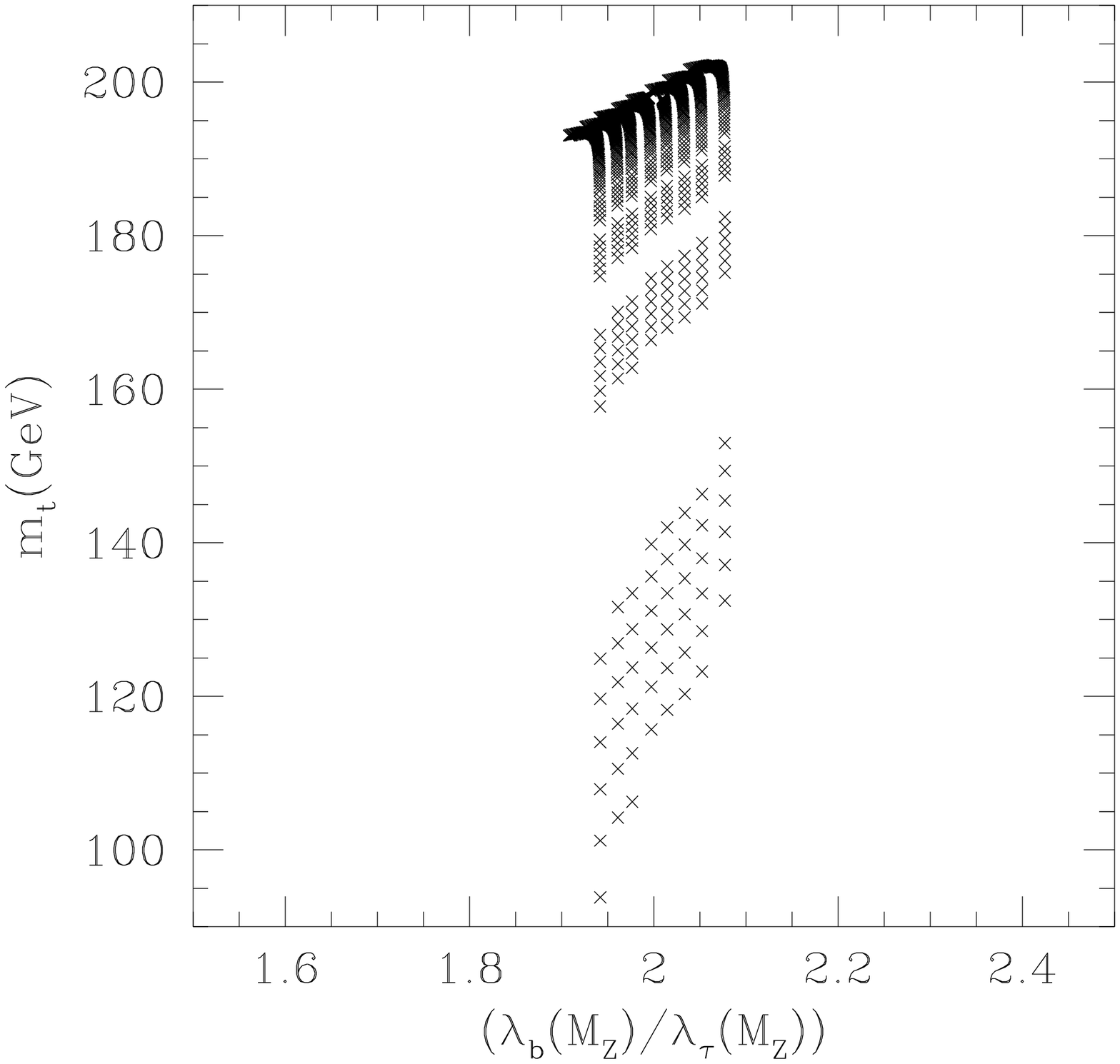}}
\nobreak
\narrower
\singlespace
\noindent
Figure 9.  A scatter plot of the physical top quark mass, $m_t(m_t)$
           versus the predicted ratio $({\lambda_b(M_Z)/\lambda_\tau(M_Z)})$.
           Each point represents a choice of parameters as in figure 2.
           It is observed that there is no strong dependence of the
           predicted top quark mass on the ratio of the Yukawa couplings.
           Thus, the Yukawa ratio at the $Z$ scale is mainly due to the
        QCD renormalization from the string scale to the weak scale while
        the predicted top quark mass mainly depend on the bottom quark
        Yukawa coupling, or alternatively on $\tan\beta$.

\medskip
\endinsert

%
\midinsert
\epsfxsize 4.5 truein
\centerline{\epsffile{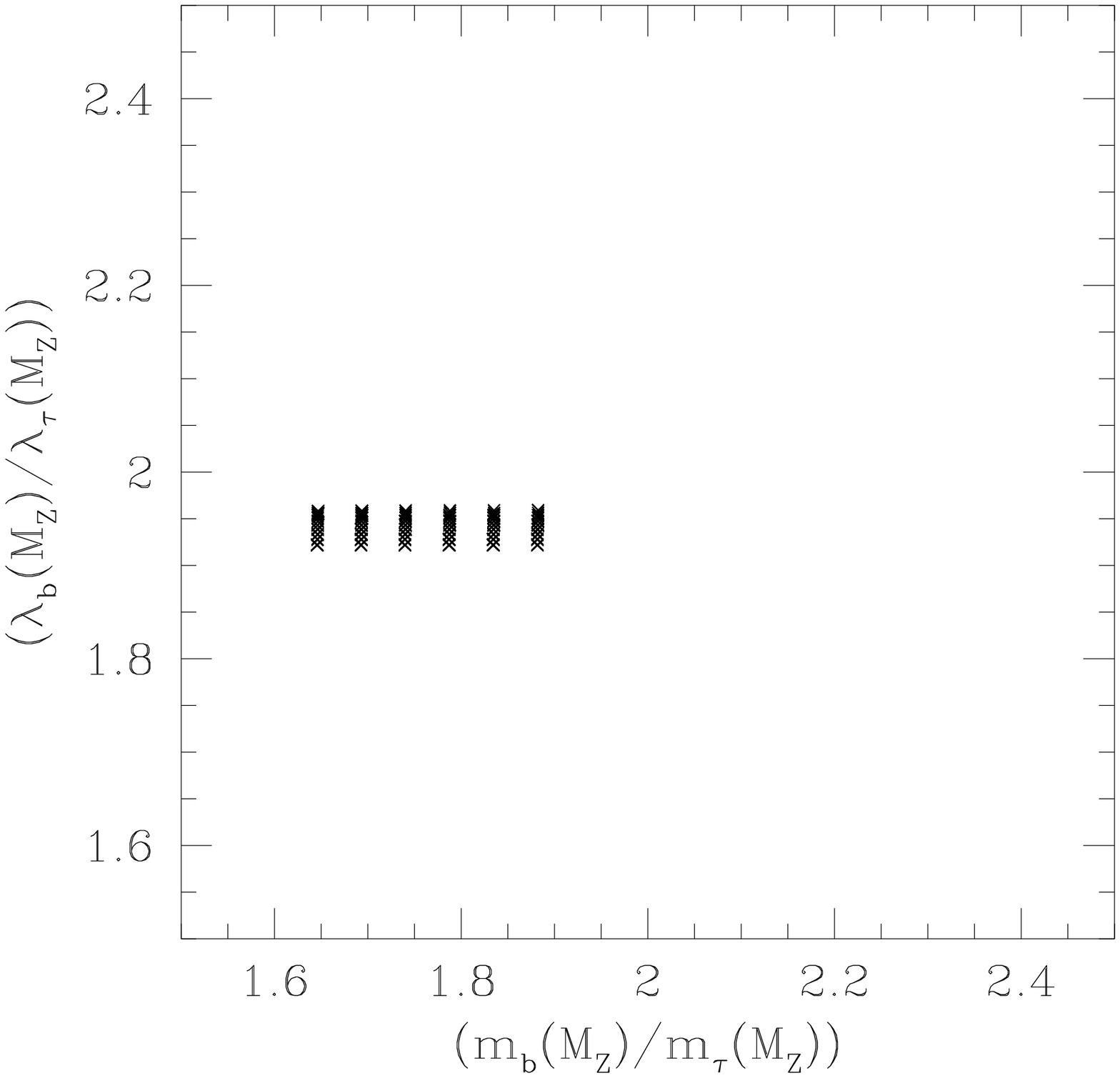}}
\nobreak
\narrower
\singlespace
\noindent
Figure 10. Scatter plot of the predicted ratio
$(\lambda_b(M_Z)/\lambda_\tau(M_Z))$ versus the experimentally
extrapolated ratio $(m_b(M_Z)/m_\tau(M_Z))$. Each point corresponds to
a point in the parameter space. The range of the SUSY breaking
parameters is given in table 3. The bottom quark mass is varied
in the range $4.1-4.7$ GeV.
\medskip
\endinsert

%
\midinsert
\epsfxsize 4.5 truein
\centerline{\epsffile{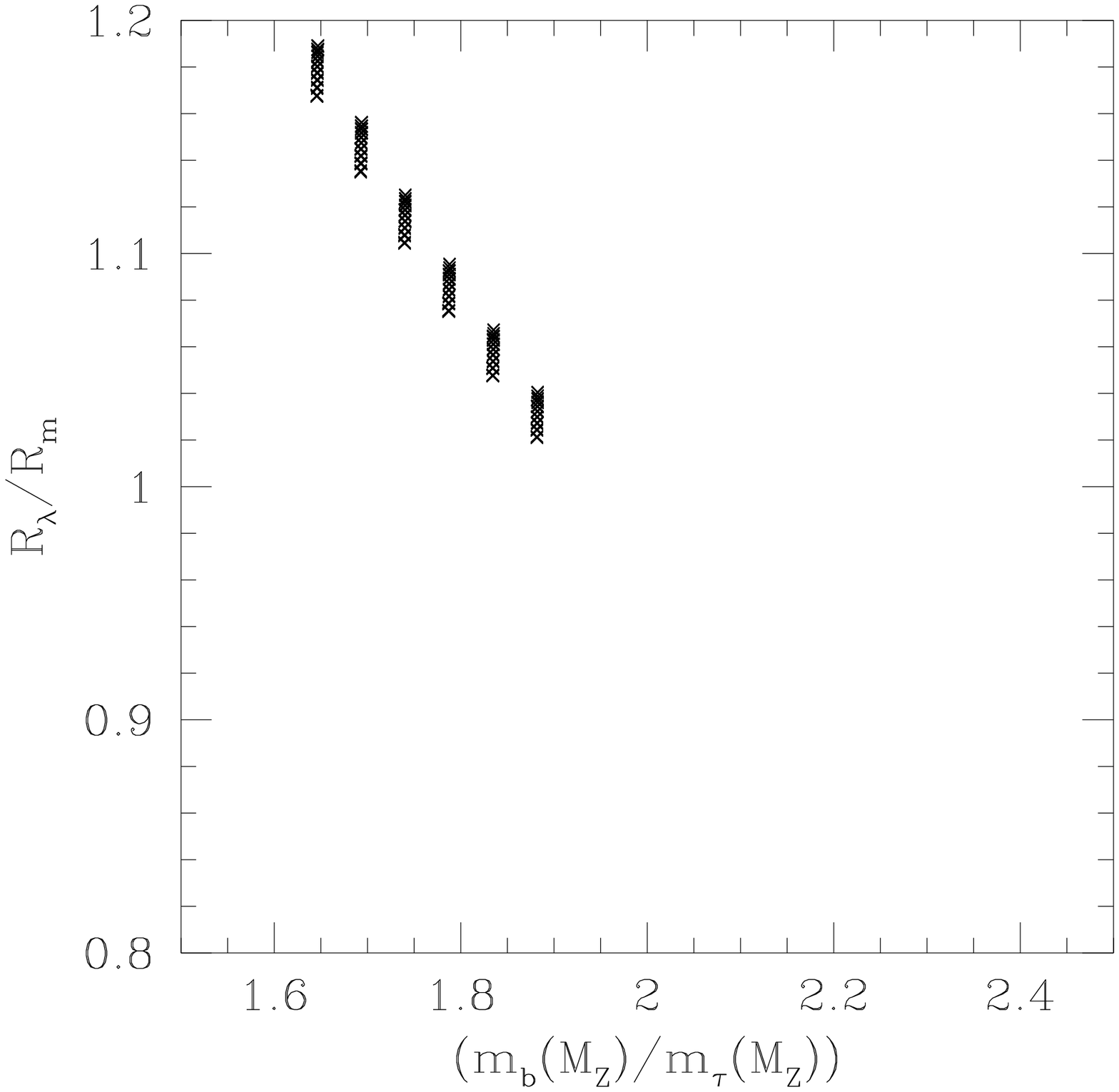}}
\nobreak
\narrower
\singlespace
\noindent
Figure 11. The ratio of $R_{\rm predicted}/R_{\rm extrapolated}$ versus
$R_{\rm extrapolated}$. Each point corresponds to a point in the
parameter space of figure 10.
\medskip
\endinsert

%

\end

\\
Title:     Calculating Fermion masses in Superstring Derived
           Standard--like Models
Author(s): Alon E. Faraggi
comments:  54 pages, phyzzx.tex, tables.tex, epsf.tex. Includes 3 tables and
           11 figures. Final version to be published in Nuclear Physics B. 
Report-no: IASSNS--95/60, UFIFT-HEP-95-29.
\\
I discuss the calculation of the heavy generation masses in the superstring 
derived standard--like models. The top quark Yukawa coupling is obtained 
from a cubic level mass term while the
bottom quark and tau lepton mass terms are obtained from nonrenormalizable
terms. The calculation of the heavy fermion Yukawa couplings is outlined in
detail in a specific toy model. The dependence of the effective bottom quark
and tau lepton Yukawa couplings on the flat directions at the string scale
is examined. The gauge and Yukawa couplings are extrapolated from the string
unification scale to low energies. Agreement with $\alpha_{\rm strong}$,
$\sin^2\theta_W$ and $\alpha_{\rm em}$ at $M_Z$ is imposed, which necessitates
the existence of intermediate matter thresholds. The needed intermediate
matter thresholds exist in the specific toy model. The effect of the
intermediate matter thresholds on the extrapolated Yukawa couplings in
studied. It is observed that the intermediate matter thresholds also help to
maintain the correct $b/\tau$ mass relation. It is found that for a large
portion of the parameter space that the LEP precision data for
$\alpha_{\rm strong}$, $\sin^2\theta_W$ and $\alpha_{\rm em}$, as well as the
top quark mass and the $b/\tau$ mass relation can all simultaneously be
consistent with the superstring derived standard--like models. Possible
corrections due to the supersymmetric mass spectrum are studied as well as the
minimization of the supersymmetric Higgs potential. It is demonstrated that
the calculated values of the Higgs VEV ratio, $\tan\beta=v_1/v_2$, can be
compatible with the minimization of the one--loop
effective Higgs potential.

\\

\bye